\documentclass[numberedappendix]{emulateapj}
\shorttitle{Mid-IR Star Formation Rate Indicator}
\shortauthors{Rujopakarn et al.}

\newcommand{\LMIPS}{\mbox{$L(24~\micron)$}}
\newcommand{\Lsun}{\mbox{$L_\odot$}}
\newcommand{\LIR}{\mbox{$L({\rm TIR})$}} 
\newcommand{\Msun}{\mbox{${\cal M}_\odot$}}
 
\newcommand{\LIRSD}{\mbox{$\Sigma_{L({\rm TIR})}$}} 
\newcommand{\SFRSD}{\mbox{$\Sigma_{\rm SFR}$}} 
\newcommand{\Td}{\mbox{T$_d$}} 

\slugcomment{Accepted for publication in The Astrophysical Journal, February 21st, 2013}
\begin{document}
\title{Mid-Infrared Determination of Total Infrared Luminosity and Star
  Formation Rates of Local and High-Redshift Galaxies}

\author{W. Rujopakarn\altaffilmark{1}, G. H. Rieke\altaffilmark{1},
  B. J. Weiner\altaffilmark{1}, P. 
  P{\'e}rez-Gonz{\'a}lez\altaffilmark{2, 1}, M. Rex\altaffilmark{1}, 
  G. L. Walth\altaffilmark{1}, J. S. Kartaltepe\altaffilmark{3}}
\altaffiltext{1}{Steward Observatory, The University of Arizona,
  Tucson, AZ 85721; wiphu@as.arizona.edu}
\altaffiltext{2}{Universidad Complutense de Madrid, Facultad de
  Ciencias F{\'i}sicas Dpto. de Astrof{\'i}sica y CC. de la
  Atm{\'o}sfera Madrid 28040, Spain} 
\altaffiltext{3}{National Optical Astronomy Observatory, 950 North
  Cherry Avenue, Tucson, AZ 85719, USA}

\begin{abstract}
We demonstrate estimating the total infrared luminosity, \LIR, and
star formation rates (SFRs) of star-forming galaxies at redshift $0 <
z < 2.8$ from single-band 24 $\micron$ observations, using local SED
templates without introducing additional free parameters. Our method
is based on characterizing the spectral energy distributions (SEDs) of
galaxies as a function of their \LIR\ surface density, which is
motivated by the indications that the majority of IR luminous
star-forming galaxies at $1 < z < 3$ have extended star-forming
regions, in contrast to the strongly nuclear concentrated,
merger-induced starbursts in local luminous and ultraluminous IR
galaxies. We validate our procedure for estimating \LIR\ by comparing
the resulting \LIR\ with those measured from far-IR observations, such
as those from {\it Herschel} in the Extended {\it Chandra} Deep Field
South (ECDFS) and {\it Hubble} Deep Field North (HDFN), as well as
\LIR\ measured from stacked far-IR observations at redshift $0 < z <
2.8$. AGNs were 
excluded using X-ray and $3.6-8.0$ $\micron$ observations, which are
generally available in deep cosmological survey fields. The Gaussian
fits to the distribution of the discrepancies between the
\LIR\ measurements from single-band 24 $\micron$ and {\it Herschel}
observations in the ECDFS and HDFN samples have $\sigma < 0.1$ dex,
with $\sim 10\%$ of 
objects disagreeing by more than 0.2 dex. Since the 24 $\micron$
estimates are based on SEDs for extended galaxies, this agreement
suggests that $\sim 90\%$ of IR  galaxies at high $z$ are indeed much
more physically extended than local counterparts of similar \LIR,
consistent with recent independent studies of the fractions of
galaxies forming stars in the {\it main-sequence} and {\it starburst}
modes, respectively. Because we have not introduced empirical
corrections to enhance these estimates, in principle, our method
should be applicable to lower luminosity galaxies. This will enable
use of the 21 $\micron$ band of the Mid-Infrared Instrument (MIRI)
on board the {\it JWST} to provide an extremely sensitive tracer of
obscured SFR in individual star-forming galaxies across the peak of
the cosmic star formation history.  
\end{abstract}

\section{INTRODUCTION}
The mid-infrared (mid-IR) is a unique window to study the evolution of  
star-forming galaxies, especially at redshift $1 < z < 3$ where the
cosmic star formation rate (SFR) peaks
\citep[e.g.,][]{HopkinsBeacom06}. At these redshifts, a majority of
star formation took place in obscured environments, where dust
reprocesses the UV photons from hot young stars into IR emission
\citep[see e.g.,][]{LeFloch05, PPG05, Dole06, Buat07}. Star-forming
galaxies at these redshifts also exhibit a large spread of extinction
values and diverse dust distribution scenarios
\citep{Rujopakarn12}. These factors pose inherent challenges for
optical and UV estimators of the SFR, which need to be complemented by
IR techniques. 

In the past decade, {\it ISO}, {\it Spitzer}, {\it WISE}, {\it AKARI},
and {\it Herschel} have allowed us to study star formation from the
local Universe out to high $z$ using mid-IR and far-IR
observations. Measurements that determine the IR luminosity, \LIR,
from these missions trace the energy absorbed from UV photons emitted 
by short-lived massive stars \citep[see, e.g.,][]{Kennicutt98}. The
most direct approach to measure \LIR\ is to use multi-band mid- and
far-IR photometry. Recent examples use {\it Herschel} to observe
star-forming galaxies at $100 - 500$ $\micron$ and fit galaxy SEDs to
measure \LIR\ \citep[e.g.,][]{Elbaz10, Rex10, Elbaz11}. The complete 
characterization of the spectral peak of the dust emission provides a
good measurement of \LIR\ and SFR. However, the inherent requirement
of multi-band photometry for far-IR SED fitting compromises this
approach for faint galaxies whose detection is limited by confusion
noise, particularly at the longer wavelengths
\citep[e.g.,][]{Condon74, Dole04}. A second approach is to use a
monochromatic (single-band) IR luminosity to trace \LIR\ and
SFR. Locally, the rest-frame single-band 24 $\micron$ luminosity has been
shown to be one of the best tracers for \LIR\ and SFR
\citep{Calzetti07, Rieke09}. Although the longer wavelengths, such as
70 $\micron$ and 160 $\micron$, are closer to the peak of the thermal
dust emission and thus reduce the size of the bolometric corrections,
they are affected more by the cold dust heated by old stars, which
increases scatter in the SFR calibration compared to estimates from 24
$\micron$ \citep{Rieke09, Kennicutt09}, particularly in less luminous
IR galaxies \citep{Calzetti10a}.

Beyond the local Universe, however, the redshifted 24 $\micron$ band
probes wavelengths containing the aromatic emissions (e.g., polycyclic 
aromatic hydrocarbons, hereafter PAH); the strongest PAH emission
complexes at 6.2, 7.7, and 8.6 $\micron$ \citep[e.g.,][]{JDSmith07}
redshift into the 24 $\micron$ band by $z \sim 2$. Although they help
boost the 24 $\micron$ flux and aid detection of galaxies at
high $z$, PAHs pose two challenges to using single-band 24 $\micron$
observations to estimate \LIR\ and SFR. First, the PAH emission is
influenced by environment (e.g., UV radiation, optical depth),
introducing a $\sim 0.2-$dex scatter to the \LIR\ and SFR estimates
\citep{Roussel01}. Second, the PAH emission appears to strengthen
intrinsically at high $z$ compared to that found in local galaxies
with the same \LIR\ \citep{Rigby08, Farrah08, Takagi10}. Therefore,
the bolometric corrections to determine \LIR\ measured from local
galaxies in the PAH wavelength  regions will not be applicable at high
$z$. Recent studies  indicate that applying the local bolometric
corrections to high-$z$ galaxies will overestimate their \LIR\ and SFR
by up to an order of magnitude. This is reported as the ``mid-IR
excess'' in recent far-IR studies using {\it Spitzer} and {\it
  Herschel} \citep[e.g.,][]{Papovich07, Elbaz10, Nordon10, Barro11,
  Nordon11}. The SED evolution causes a {\it systematic bias} that
must be taken into account to use 24 $\micron$ observations as
\LIR\ and SFR indicators beyond the local Universe. 

Out to $z \sim 3$, the {\it Spitzer} 24 $\micron$ observations probe
weaker SFR at any given redshift than is possible with the far-IR 
bands \citep[see Figure 4 of][]{Elbaz11}. For {\it JWST}, 21 $\micron$
is the longest wavelength band suitable for deep cosmological
surveys. Therefore, our understanding of star-forming  galaxies at
high $z$ will depend critically on mid-IR SFR indicators, where the
current state-of-the-art prescription to estimate \LIR\ from
single-band 24 $\micron$ observations still presents a $0.4-$dex
scatter \citep{Nordon11}. Additionally, the {\it Spitzer} 24 $\micron$
data are already available from deep legacy surveys (e.g., GOODS,
FIDEL, COSMOS, SpUDS), as well as from large-area surveys (e.g., SWIRE
and the Bo\"{o}tes field) that will not be fully surveyed by current
facilities to the same depth in terms of SFR. Exploration of means to
reduce the current $0.4-$dex scatter in \LIR\ estimates will allow
utilization of the mid-IR observations to the fullest extent in the
upcoming decade. 

In this paper, we apply the results from our previous study,
\citet{Rujopakarn11}, to take into account the SED evolution of
star-forming galaxies and refine the 24 $\micron$ \LIR\ and SFR
indicators. \citet{Rujopakarn11} find that the IR luminosity surface 
density, \LIRSD, affords an accurate description of the SED and
subsequently allows accurate bolometric corrections out to high $z$,
specifically out to $z = 2.8$, the farthest redshift where the {\it
  Spitzer} 24 $\micron$ band traces predominantly PAH emissions. The
measurement of \LIRSD\ requires high-resolution imaging of the 
star-forming regions, which is only available for a small number of
galaxies. We use these galaxies as a tool to construct a simple
formula to estimate \LIR\ and SFR without measuring \LIRSD\ for
individual galaxies and using only single-band 24 $\micron$
flux and redshift measurements, and then compare the resulting
\LIR\ estimates with results using far-IR data. 

This paper is organized as follows. We discuss the evolution of the
SEDs of star-forming galaxies and the use of \LIRSD\ as a guide to
select the appropriate SED for high-$z$ galaxies in Section
\ref{sec:sedevol}. The derivation of the single-band
\LIR\ indicator is described in Section \ref{sec:indicator}. We
then compare the new \LIR\ indicator with other independent
measurements in Section \ref{sec:indicator_test} and present the
\LIR\ SFR indicator in Section \ref{sec:final_SFR}. Lastly, we discuss
the validity of rest-frame $8-24$ $\micron$ as a tracer of \LIR\ and
SFR, along with implications of our results in Section
\ref{sec:discuss}. Throughout this paper, we assume a $\Lambda$CDM
cosmology with  $\Omega_m = 0.3$, $\Omega_{\Lambda} = 0.7$, and $H_0 =
70$~km~s$^{-1}$Mpc$^{-1}$. We follow the convention of \citet{Rieke09}
and adopt the definition of \LIR\ of \citet{SandersMirabel96}; these
studies defined \LIR\ slightly differently (i.e., 
$5-1000$ $\micron$ vs. $8 - 1000$ $\micron$), but the resulting
\LIR\ values are closely consistent.  We will refer to galaxies with
\LIR\ in the range of $10^{11}-10^{12}$ and those with
\LIR\ $>10^{12}$ \Lsun\ as Luminous Infrared Galaxies (LIRGs) and
Ultraluminous Infrared Galaxies (ULIRGs), respectively, or
collectively as U/LIRGs.\\

%
%
%
%

\section{QUANTIFYING THE IR SED EVOLUTION}\label{sec:sedevol} 
In this section, we will discuss the challenges in using {\it Spitzer}
observations at 24 $\micron$ to estimate \LIR\ and SFR. Major issues
are to determine an overall approach for using \LIRSD\ as an
indicator of a galaxy SED (Section \ref{sec:sedevol_intro}), and the
choice of SED library (Section \ref{sec:sedevol_sedchoice}).

\subsection{\LIRSD\ as an Indicator of SEDs for Star-Forming
  Galaxies}\label{sec:sedevol_intro} 
Until now, most estimations of the \LIR\ and SFR of galaxies using
single-band IR observations rely on an assumption that the SED of a
star-forming galaxy does not evolve with redshift, i.e., that
bolometric corrections measured from local SED templates can be
applied to high-$z$ galaxies. However, local IR 
galaxies, from which we construct the SED libraries, comprise an
inhomogeneous population of both normal star-forming galaxies
(quiescent disks) and those with nuclear star-formation induced by 
galaxy interactions. The contribution of interaction-induced star
formation increases with \LIR; locally, theoretical studies suggest
that ULIRGs must be dominated by interaction-induced starbursts to
achieve their \LIR\ outputs \citep{Hopkins10}, which is
consistent with observations \citep[e.g.,][]{SandersMirabel96,
  Veilleux02}. The application of local bolometric corrections to
high-$z$ galaxies thus carries an implicit assumption that high-$z$
star-forming  galaxies are likewise dominated by interaction-induced
starbursts. 

Observations have shown, however, that actively star-forming galaxies
at high $z$ are different from their local counterparts in at least
three major aspects. First, the IR SEDs at high $z$ exhibit colder
far-IR dust temperatures, \Td, than local galaxies at fixed
\LIR\ \citep{Pope06, Symeonidis09, Muzzin10}. These galaxies have dust
temperatures similar to local galaxies with lower \LIR. Second is the
aforementioned evolution of the strength of the aromatic features,
which grow stronger at high $z$ at a fixed \LIR. Quantitatively,
for instance, \citet{Rigby08} found Abell 2218a, a galaxy with
\LIR\ of $10^{11.9}$ \Lsun\ at $z = 2.5$, to exhibit aromatic emission
features virtually identical to those of a local galaxy with an order
of magnitude lower \LIR. 

Third, \citet{Rujopakarn11} found that the
diameters of the galactic-wide star-forming regions in high-$z$ 
U/LIRGs are $10-30\times$ larger than those of local U/LIRGs at
the same \LIR. The size of the star-forming regions found in high-$z$
U/LIRGs is similar to that of local normal star-forming galaxies
(sub-LIRG), $4-10$ kpc in diameter, but with \SFRSD\ scaled up by a
factor of $100 - 1000\times$ \citep[Figures 3 and 4
  of][]{Rujopakarn11}. This structural similarity is a manifestation
of the ``main sequence'' of star-forming galaxies, originally defined
as a sequence of galaxies in stellar mass vs. SFR space
\citep{Noeske07}, where star-formation occurs at a relatively steady
rate rather than in bursts, and in disks or clumps, rather than in
merger  nuclei \citep[e.g.,][]{Genzel10, Tacconi10, Narayanan10,
  Daddi10, Elbaz10, Elbaz11}. In its simplest form, this picture
suggests that these galaxies differ among themselves (and from local
lower luminosity star-forming galaxies) primarily in the SFR surface
density, \SFRSD, and hence in IR luminosity surface density, \LIRSD.

\begin{figure}
\figurenum{1}
\epsscale{1.2}
\plotone{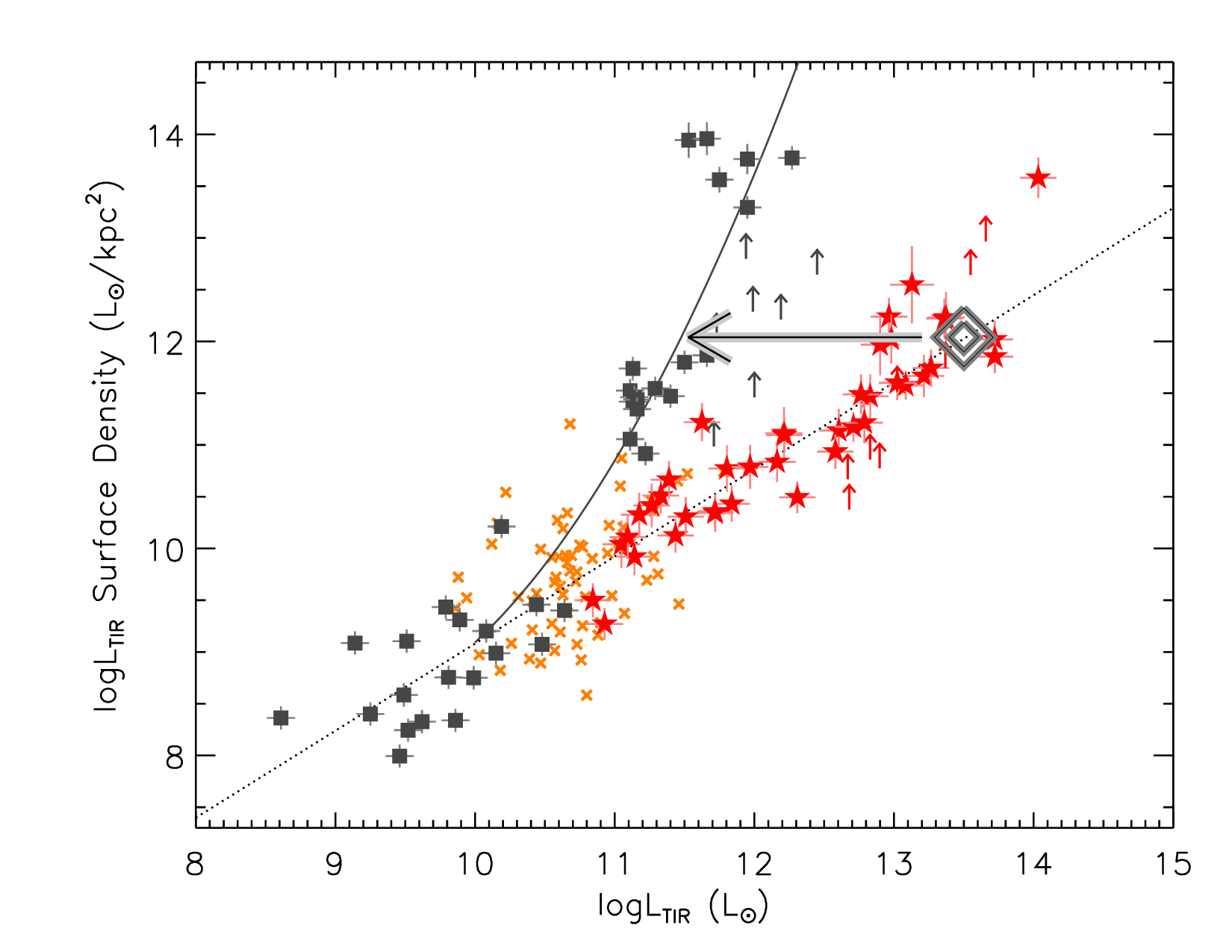}
\caption{The relationships between the \LIR\ and \LIR\ surface
  density, \LIRSD, differ for the local and non-local galaxies (grey
  squares and red stars, respectively). Local galaxies from
  \citet{Chanial07} are shown for comparison as orange crosses. The
  upward 
  arrows represent points derived from CO size measurements, which
  could be systematically more extended than the other size tracers
  employed \citep{Rujopakarn11} and thus are taken as lower limits in
  \LIRSD. The solid and dashed lines are the fits for local and
  high-$z$ galaxies (excluding the CO lower limits),
  respectively. \citet{Rujopakarn11} show that the SEDs of galaxies
  are to first order indicated solely by their \LIRSD. For 
  example, a high-$z$ galaxy with \LIR\ of $10^{13.5}$ \Lsun will have 
  aromatic emissions and absorption features consistent with a local
  galaxy with the same \LIRSD, indicated by the arrow, in this case a
  local galaxy with \LIR\ of $\sim 10^{11.5}$ \Lsun\ (see Section
  \ref{sec:indicator}). This behavior, plus the tendency of the
  great majority of high-$z$ galaxies to lie near the main sequence, 
  allows a simple approach to associating the appropriate SED with
  typical high-$z$ galaxies.}
\label{lir_lirsd}
\end{figure}

In Figure \ref{lir_lirsd}, adapted from Figure 4 of
\citet{Rujopakarn11}, we show \LIRSD\ as a function of \LIR\ along
with trend lines describing the relationship in each population, one
for local galaxies with \LIR\ above $10^{11}$ \Lsun\ and another for
high-$z$ galaxies. The main sequence can be drawn in this figure
from the local normal star-forming galaxies (sub-LIRG galaxies) onto
the high-$z$ star-forming galaxies. Both the local and
high-$z$ relationships agree below \LIR\ $\sim 10^{11}$ \Lsun. The
trends diverge above the LIRG threshold, where the local galaxy sample
starts to be dominated by galaxies harboring interaction-induced star
formation (see also \citet{Totani11} for the local relationship).

\citet{Rujopakarn11} estimated \LIRSD, defined by \LIRSD\ $=$ 
  \LIR/$A$, by measuring the area, $A$, of the IR-emitting
  region using radio continuum (e.g., 1.4 GHz), Paschen-$\alpha$,
  and 24 $\micron$ observations. Although the emission physics of
  these size measures is different, they trace 
a consistent physical extent of the star-forming regions and
are similarly insensitive to contributions from old stellar
populations. Paschen-$\alpha$ and 24 $\micron$ are the two best IR
tracers of SFR and are emitted from approximately the same
spatial extent. Also, \citet{Chanial07} found that radio continuum
(e.g., 1.4 GHz) and far-IR size measurements are consistent within
15\%. For the purpose of determining the surface area of 
star formation in individual galaxies, we will thus use these
three size tracers interchangeably. Among these indicators, only the
interferometric observation of the radio synchrotron continuum can
resolve individual galaxies at high-$z$. The main sample of
high-$z$ galaxies in the Figure \ref{lir_lirsd} was therefore from the
\citet{Muxlow05} 1.4 GHz VLA+MERLIN survey of the {\it Hubble} Deep
Field with angular resolution of $0\farcs2 - 0\farcs5$. This is a
blind survey that has resolved all of the 92 detected
galaxies, which assures that the extended physical
sizes at high $z$ are not due to selection bias. The Muxlow et
al. sample is augmented by radio observations of sub-mm galaxies from
the literature with similarly high angular resolutions. The samples of
galaxies with physical size measurements is tabulated in the online
material accompanying this paper for interested readers. 

Following the \citet{Rujopakarn11} example, the use of \LIRSD\ as an
indicator of SED characteristics can be illustrated by envisioning a
horizontal line at a fixed \LIRSD\ in Figure \ref{lir_lirsd}. As an
example, if we assume the $z = 2.5$ Abell 2218a with \LIR\ of
$10^{11.9}$ \Lsun, observed by \citet{Rigby08}, is located on the
main-sequence, we would expect it to exhibit spectral characteristics
of a galaxy on the local sequence with \LIR\ of $\sim10^{10.9}$
\Lsun\ that has the same \LIRSD. \citet{Rujopakarn11} further
demonstrate the ability of \LIRSD\ to predict SED behavior by 
predicting the 24 $\micron$-to-1.4 GHz flux ratios, which are
sensitive to PAH features at $z \sim 2$, consistently with the
observed ratios (their Figure 5), as well as 
matching the average observed aromatic spectrum of high-$z$ ULIRGs to
a local SED template (their Figure 7). The assumption that \LIRSD\ is
the dominant parameter controlling the average SED thus provides a
tool to assign an appropriate local SED template to represent 
the SED of high-$z$ galaxies, a result we will use to construct an
indicator for \LIR\ and SFR in this work.

The \LIRSD\ method is physically motivated because the PAH
emission emerges from the outer surfaces of the photodissociation
regions \citep[PDRs; e.g.,][]{Tielens08}, so the surface density of
the SFR is a controlling factor for the PAH-emitting surface area and
thus the PAH emission strength. Additionally, \citet{CM05} modeled the
radiative transfer in centrally heated dusty 
sources and found a general relation among the luminosity-to-mass
ratio, the surface density, and the shape of the SED. They predict 
that the behavior of SED templates for local galaxies might not extend
to high $z$. Extended sizes and colder SEDs were also shown to be
connected through a \Td-\LIRSD\ relation explored by
\citet{Chanial07}, who theoretically described the IR-emitting region
of a star-forming galaxy as an isothermal cloud optically thick to
optical wavelengths and optically thin in the IR. \citet{Chanial07}
show that \LIR, \Td, and the extent of the IR-emitting region are
related by a fundamental plane; the \LIR-\Td\ relation
\citep[e.g.,][]{Soifer87} is a manifestation thereof. 

\subsection{The Local Reference for the Galaxy
  SEDs}\label{sec:sedevol_sedchoice}   
The SED libraries commonly used to represent the spectra of 
star-forming galaxies are those of, e.g., \citet{CharyElbaz01},
\citet{DaleHelou02}, and \citet{Rieke09}. The former two libraries
were based on spectra taken by {\it ISO}, preceding the availability
of full spectral coverage at $6 - 24$ $\micron$ (that the 24 
$\micron$ band probes at redshift $z = 2.8 - 0$). The recent 
infrared spectrograph (IRS) observations from {\it Spitzer}, covering
$5.2 - 38$ $\micron$, show  that the \citet{CharyElbaz01} SED
templates (hereafter, CE01) have suppressed aromatic features at very
high luminosity \citep{CharyPope10} and do not sufficiently take into
account the strong silicate absorption features at 10 $\micron$,
resulting in weaker aromatic bands but stronger net emission in the
PAH-region comparing to observed galaxies \citep{Armus07,
  Rieke09}. The \citet{DaleHelou02} SED library is optimized to
describe moderately-luminous local star-forming galaxies with \LIR\ $<
10^{11}$ \Lsun\ and hence the lack of silicate absorption in these
templates becomes more significant at higher \LIR\ with larger typical
extinction.

\citet{Rieke09} developed SED templates (hereafter, R09), which we
adopt as the SED reference for local galaxies, separately for normal
star-forming galaxies (sub-LIRG) and U/LIRGs to provide a
self-consistent SED library covering \LIR\ of $10^{9.75}$ to
$10^{13.00}$ \Lsun. The 
U/LIRGs SED templates were constructed from a sample of 11 local LIRGs 
and ULIRGs with high quality {\it Spitzer} IRS spectra as well as
photometric data covering optical to radio wavelengths (see Figures 1,
2, and 3 of \citet{Rieke09}). These U/LIRGs were chosen such that
their IR emission is dominated by star-forming activity. Construction 
of the SED library is done in two steps. First, these 11 galaxies were
used as a basis to assemble 11 archetypal SED templates spanning 0.4
$\micron$ to 30 cm wavelength. Their IRS spectra ($5-38$ $\micron$)
were joined to the photospheric emission and far-IR dust emission
components in a series of tests to ensure both spectral continuity and 
appropriate flux calibration. Second, these archetypal templates
were combined with different weights to produce the final averaged SED
templates. The template weights were optimized by matching synthesized
IR colors from the combined template to the average IR colors of
observed galaxies as a function of \LIR\ from IRAC and the {\it IRAS}
RBGS \citep{Sanders03}. The R09 
library construction was extended to star-forming galaxies with
sub-LIRG \LIR\ by combining the \citet{DaleHelou02} SED library with
the mid-IR spectral library based on IRS observations from
\citet{JDSmith07} using the same IR color fitting technique as in the
U/LIRG template construction. The use of IR color to help guide the
combination of archetypal templates (i.e., fitting $25/8$ $\micron$,
$25/12$ $\micron$, and $60/25$ $\micron$ colors simultaneously) helps
ensure that the final templates represent the average properties of
real galaxies even though they are constructed from a limited
sample. The R09 SED library construction is described in detail in 
the Appendix of \citet{Rieke09}. We adopt these templates for this
paper. \\

%
%
%
%

\section{A MID-IR ESTIMATOR FOR \LIR\ AT $0 < z <
  2.8$}\label{sec:indicator}  
The construction of a 24 $\micron$ SFR indicator has two steps: (1) 
construct an \LIR\ estimator using 24 $\micron$ flux and redshift, and then
(2) derive a relationship between \LIR\ and SFR. The first step
would appear to require measuring \LIRSD\ for each galaxy to determine
the appropriate SED template. However, the relatively small scatter of
the high-$z$ galaxies around the main sequence in Figure
\ref{lir_lirsd} suggests that adequate SED template matching can be
achieved by assuming a typical surface area. In this Section, we will
use the sample with \LIRSD\ measurements in Figure \ref{lir_lirsd} to
construct a formula that can be applied to estimate SFRs in the
absence of the physical size measurements. 


\citet{Rieke09} first assign a value of \LIR\ to each SED template and
then determine the corresponding SFR as well as the monochromatic
luminosity through the desired bandpasses. For the 24 $\micron$ band,
in particular, the assignment of the 24 $\micron$ luminosity of a
template follows the ratio of \LMIPS-to-\LIR\ based on the local {\it 
  IRAS} data \citep[see Fig. 8, Fig. 15, and formula A6
  of][]{Rieke09}. Once each template has an associated 24 $\micron$
luminosity, it is possible to calculate the k-corrections and
subsequently the expected monochromatic flux for each template as a
function of redshift. For $L_{\nu} \propto \nu f_{\nu} $, the
k-correction, ${\rm K}_{{\rm corr}}(z)$, and the flux, $f_\nu(24 \mu
m)$, are related by  

\begin{equation}
\label{eq1}
{\rm K_{corr}}(z) = (1+z) \frac{f_{\nu}(\nu=(1+z)\nu_{\rm
    {obs}})}{f_\nu(24 \mu {\rm m})} 
\end{equation}

\begin{equation}
\label{eq2}
4 \pi {D_L}^2 f_{\nu, {\rm obs}} = \frac{L_{\nu, {\rm rest}}(24 \mu
  m)}{\nu_{24}}{\rm K_{corr}}(z)
\end{equation}

\noindent where $D_L$ is the luminosity distance for an object at
redshift  
$z$. \noindent The $4 \pi{D_L}^2 f_{24, {\rm obs}}$ for this
equation has the units of Jy cm$^2$. The relationship between the
observed monochromatic flux from each template and the template's
monochromatic luminosity at a given redshift is approximately linear,
which allows for a linear fit at each redshift
(i.e., for a redshift grid) to determine a set of coefficients that
convert the observed flux at a given redshift to the luminosity or
any quantity associated with the template (e.g., \LIR\ and SFR). For
example, a relationship between \LIR\ and the observed 24 $\micron$ 
flux, $f_{24,{\rm obs}}$, has the form 

\begin{equation}
\label{eq3}
{\rm log} \LIR_{z = 0} = A(z) + B(z) \left[ {\rm log} (4 \pi
  {D_L}^2 f_{24, {\rm obs}}) - C \right],
\end{equation}

\noindent where $A(z)$ and $B(z)$ are the intercept and the slope from
the linear fit, respectively, $C$ is a zero-point to reduce
covariance in the fit parameters, and \LIR$_{z = 0}$ refers to the
\LIR\ associated with each of the R09 SED templates. \citet{Rieke09} use
this method to tabulate the coefficients to convert the monochromatic
fluxes in various bands to the SFR of galaxies.

This formalism successfully estimates \LIR\ and SFR for local
galaxies. However, a modification is needed to apply the formalism at
high $z$ \citep{Rieke09}. If we assume that IR star-forming galaxies
beyond the local Universe are in the main sequence (we discuss the
validity and applicability of this assumption in Section
\ref{sec:discuss_sfmodes}), it is possible to use the corresponding
\LIRSD\ as a guide to estimate \LIR\ and SFR at high $z$ without the
need for \SFRSD\ measurements of individual galaxies. Following 
\citet{Rujopakarn11}, an appropriate choice of SED to use for
calculations of the k-correction and bolometric correction beyond the
local Universe is 
the one corresponding to the same \LIRSD\ locally. This is equivalent
to reassigning the \LIR\ associated with each of the R09 SED templates
to new values determined by the ratio of luminosities with equal
\LIRSD\ on the local U/LIRG trend to those on the main sequence. We
will refer to this ratio as the {\it stretching factor}, $S_i$, for
each SED template. From Figure \ref{lir_lirsd}, $S_i$ is nearly
negligible at \LIR$_{z=0} = 10^{11}$ \Lsun\ and reaches three 
orders of magnitude at \LIR$_{z=0}$ of $10^{14}$ \Lsun. To determine
$S_i$ quantitatively, we parameterize the relationship by fitting a
parabola to the local U/LIRG relation (excluding the
lower limits of \LIRSD\ from CO observations) and a linear fit to the
main sequence, and then take the ratio of \LIR\ on the main sequence
fit to the R09 template \LIR\ at the same \LIRSD\ (see Figure
\ref{lir_lirsd}; the $S_i$ values are tabulated in Table
\ref{table_si}). Since we effectively increase the luminosity of each
template by a factor of $S_i$, the observed flux will also be
increased by the same factor. Equation \ref{eq3} can thus be rewritten
as 
\begin{eqnarray}
\label{eq4}
{\rm log\LIR_{{\rm new}}} = {\rm log\left[S_i\LIR_{z = 0}\right]} \nonumber \\ 
= A'(z) + B'(z) \left[ {\rm log} (4 \pi {D_L}^2 S_if_{24, {\rm obs}})
  - C' \right]
\end{eqnarray}

The new set of coefficients, $A'(z)$, $B'(z)$, and $C'$, can be
determined by re-fitting equation \ref{eq4}. We have limited the
fitting range to only encompass the ``stretched'' luminosity of
log\LIR$_{{\rm new}}$ $< 14$ because the stretched luminosities at
large $S_i$ are far greater than the luminosity range occupied by real
galaxies (the $10^{14}$ \Lsun\ cut off is chosen because we expect
this to be approximately at luminosity limit of star-forming
galaxies). For a given template, the shape as a function of redshift
remains the same.  The effect of the $S_i$ is to change the spacing
between templates.  The relationship between luminosity and observed
flux remains well approximated as linear, with residuals $<0.05$ dex.

\begin{center}
\begin{deluxetable}{ccc}
\tablewidth{0pt}
\tablecaption{Stretching Factors and \LIR$_{{\rm new}}$ associated to
  \citet{Rieke09} SED Templates}
\tablehead{log\LIR$_{z = 0}$\tablenotemark{$\dagger$} & log$S_i$ &
  log\LIR$_{{\rm new}}$} 
\startdata
 9.75 &   -0.118 &   9.63\\ 
10.00 &    0.013 &  10.01\\
10.25 &    0.173 &  10.42\\
10.50 &    0.408 &  10.91\\
10.75 &    0.717 &  11.47\\
11.00 &    1.101 &  12.10\\
11.25 &    1.560 &  12.81\\
11.50 &    2.095 &  13.59\\
11.75 &    2.704 &  14.45\\
12.00 &    3.388 &  15.39
\enddata
\tablenotetext{$\dagger$}{The \citet{Rieke09} SED templates with
  \LIR$_{z = 0} > 10^{12}$ \Lsun\ are omitted because their
  \LIR$_{{\rm new}}$ are higher than luminosities of observed
  galaxies. These SED template shapes are thus unlikely to represent
  real galaxies at high $z$.}
\tablecomments{Col. (1) Original \LIR\ associated to each of the
  \citet{Rieke09} 
  SED templates; Col. (2) stretch factors from the fit to the
  \LIR-\LIRSD\ relationship of local galaxies (Section
  \ref{sec:indicator}); Col. (3) resulting \LIR$_{{\rm new}}$
  associated to SED template shape from the fit.}
\label{table_si}
\end{deluxetable}
\end{center}

\begin{center}
\begin{deluxetable}{ccc}
\tablewidth{0pt}
\tablecaption{Coefficients of the Fits for the Relation Between {\it
    Spitzer} 24 $\micron$ Flux and the \LIR}
\tablehead{$z$ & $A'(z)$ & $B'(z)$}
\startdata
0.0 &    2.656 &    0.975 \\
0.2 &    2.350 &    1.020 \\
0.4 &    2.060 &    1.056 \\
0.6 &    2.012 &    1.065 \\
0.8 &    1.659 &    1.094 \\
1.0 &    1.296 &    1.129 \\
1.2 &    1.137 &    1.159 \\
1.4 &    1.039 &    1.179 \\
1.6 &    1.015 &    1.165 \\
1.8 &    0.934 &    1.149 \\
2.0 &    0.922 &    1.145 \\
2.2 &    0.896 &    1.149 \\
2.4 &    0.837 &    1.158 \\
2.6 &    0.768 &    1.175 \\
2.8 &    0.655 &    1.198
\enddata
\tablecomments{$A'(z)$ and $B'(z)$ are the coefficients for equation
  \ref{eq5}.}  
\label{table_azbz}
\end{deluxetable}
\end{center}

\begin{figure}
\figurenum{2}
\epsscale{1.2}
\plotone{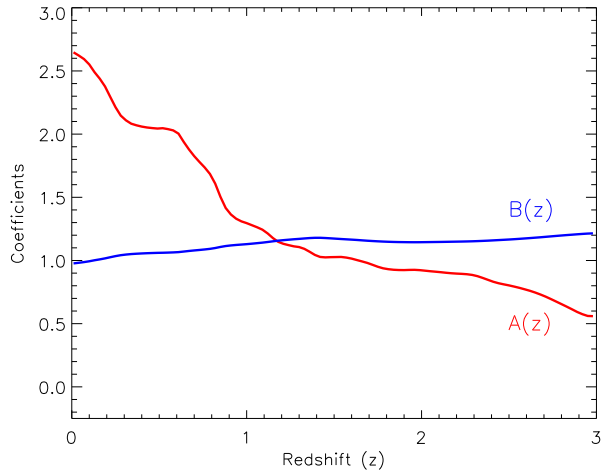}
\caption{The coefficients of the fits for a relationship between the
  observed {\it Spitzer} 24 $\micron$ flux and the total IR luminosity
  (Equation \ref{eq5}) as a function of redshift. The coefficients 
  $A'(z)$ and $B'(z)$ are shown by the red and blue
  lines. These coefficients are tabulated in Table \ref{table_azbz}.\\}
\label{azbz}
\end{figure}

The fitting coefficients, $A'(z)$ and $B'(z)$, to relate the observed
{\it Spitzer} 24 $\micron$ flux to \LIR\ as a function of redshift, are
shown in Figure \ref{azbz} (as dotted lines) and tabulated in Table
\ref{table_azbz}. The relation is \begin{equation}
\label{eq5}{\rm log\LIR_{{\rm new}}} = A'(z) + B'(z) \left[ {\rm log}
  (4 \pi {D_L}^2 f_{24, {\rm obs}}) - 45 \right], 
\end{equation}

Apart from the 24 $\micron$ single-band \LIR\ indicator, we also use
the local \LIR-\SFRSD\ relationship to derive the $A(z)$ and $B(z)$
coefficients to estimate \LIR\ from the 70  $\micron$ observed flux
(e.g., {\it Spitzer} MIPS or {\it Herschel} PACS). The resulting
coefficients yield single-band 70 $\micron$ \LIR\ consistent with
the values from the original \citet{Rieke09} estimator at \LIR\ $<
10^{11}$ \Lsun\ and overestimate the \citet{Rieke09} \LIR\ by $< 0.15$
dex at higher luminosities. We do not anticipate strong evolution of
the SED in the rest-frame wavelength range probed by the observed 70
$\micron$ band at redshift z = $0 - 2.8$ ($70 - 18.4$ $\micron$)
because it is still in the dominant blackbody
emission peak, unlike the 24 $\micron$ band. Thus the \citet{Rieke09}
coefficients can be used to estimate \LIR\ and SFR from 70 $\micron$
observations.\\

%
%
%
%

\section{TESTING THE NEW 24 $\micron$
  \LIR\ INDICATOR}\label{sec:indicator_test}  
We test the new 24 $\micron$ indicator on seven samples of galaxies
with far-IR \LIR\ measurements. Five of these samples are of
individual galaxies; two are stacked multi-band far-IR photometry. In
Section \ref{sec:indicator_testindiv}, we test the indicator on
individual galaxies with accurate far-IR photometry from {\it
  Herschel} to establish that the method can successfully estimate
\LIR\ of star-forming galaxies without systematic biases, as well as
to quantify the scatter of \LIR\ estimates for individual
galaxies. Individual gravitationally lensed galaxies at $1.0 < z <
2.7$ with far-IR and sub-mm \LIR\ measurements are also introduced in
this section to test the indicator down to the LIRG regime at $z
\gtrsim 
2$. In Section \ref{sec:indicator_teststat}, we test the
applicability of the indicator using averaged SEDs (i.e., stacked) of a
wide range of star-forming galaxies in cosmological surveys at $0 < z
< 2.8$. Additionally, we verify
the uncertainty estimate of the \LIR\ values using an independent
sample of individual  galaxies that have {\it Spitzer} far-IR
observations and photometric redshifts in Section
\ref{sec:indicator_testcosmos}.

\subsection{Testing the 24 $\micron$ \LIR\ Indicator with Far-IR
  \LIR\ Measurements for Individual
  Galaxies}\label{sec:indicator_testindiv}
The individual galaxies we use to verify that the new indicator
predicts \LIR\ consistent with that from far-IR SED fitting are from
the following: (1) {\it Herschel} SPIRE observations of galaxies in
the Extended {\it Chandra} Deep Field South (ECDFS) with spectroscopic
redshift measurements; (2) {\it Herschel} PACS and SPIRE observations
of galaxies in the {\it Hubble} Deep Field North (HDFN) with
spectroscopic redshift measurements, augmented by photometric
redshifts; (3) the {\it Herschel} Lensing Survey 
\citep[HLS;][]{Egami10, Rex10} sample at $0.4 < z < 2.7$; (4) 24
$\micron$-bright lensed galaxies at $1 < z < 2.7$ studied by
\citet{Rujopakarn12}. These far-IR data sets provide photometric
measurements near the peak of dust emission spectrum that can be used
as fiducial \LIR\ estimates, in addition to their multiwavelength
ancillary data for AGN identification. A range of $10^{11} <
\LIR\ 10^{13} < $ \Lsun\ is represented. 

\subsubsection{Extended {\it Chandra} Deep Field South Sample}
For the ECDFS sample, we processed the archival {\it Herschel} SPIRE
observations at 250, 350, and 500 $\micron$, following the method
described in \citet{Kennicutt11} to produce the final maps in each
band using HIPE  release 8.1 (SPIRE pipeline version 8.0.3287). The
original observations were from the HerMES survey (PI:
S. Oliver). Flux measurements on these maps were done with PSF
photometry using the DAOPHOT software \citep{Stetson87} on the 24
$\micron$ prior positions from the FIDEL survey (C. Papovich, private
communication) with no re-centering allowed. We measure the positional
offsets between the prior coordinates and those of the SPIRE maps by
stacking SPIRE sources and fit a 2D Gaussian to measure the offset of
the resulting centroid of the stacked PSF from the priors. These
offsets, $\alpha = -3\farcs3$, $\delta = 0\farcs8$, were applied to
the maps before performing PSF photometry. The large, $18'' - 36''$,
beams of {\it Herschel} SPIRE presents blending issues: fluxes from
nearby objects can contribute  artificially to the objects of
interest. We developed a criterion based on the 24 $\micron$ prior
catalog to exclude the potentially-blended objects in an unbiased
manner. This method is described in Appendix \ref{sec:rejectblended}. 

AGNs are excluded from this test by excluding galaxies with L$_{\rm
  X}[0.5-8.0~{\rm keV}] > 10^{42}$ erg/s and those exhibiting IR
power-law SED \citep{Donley12}, based on the {\it Chandra} X-ray and
{\it Spitzer}/IRAC catalogs from \citet{Lehmer05} and \citet{Damen11}, 
respectively. The requirement of these ancillary data to exclude AGNs
makes our method most useful where deep 24 $\micron$ observations are
accompanied by those of deep X-ray and/or mid-IR (e.g., {\it
  Spitzer}/IRAC). This is presently the case in every major
cosmological deep field, and future deep observational programs will
likely take place in these same fields (e.g., ECDFS, GOODS-N, EGS, COSMOS,
UDS). Thus, future investigations will be able to exclude AGN much as
we have.

\begin{center}
\begin{figure}[ht]
\figurenum{3}
\epsscale{1.2}
\plotone{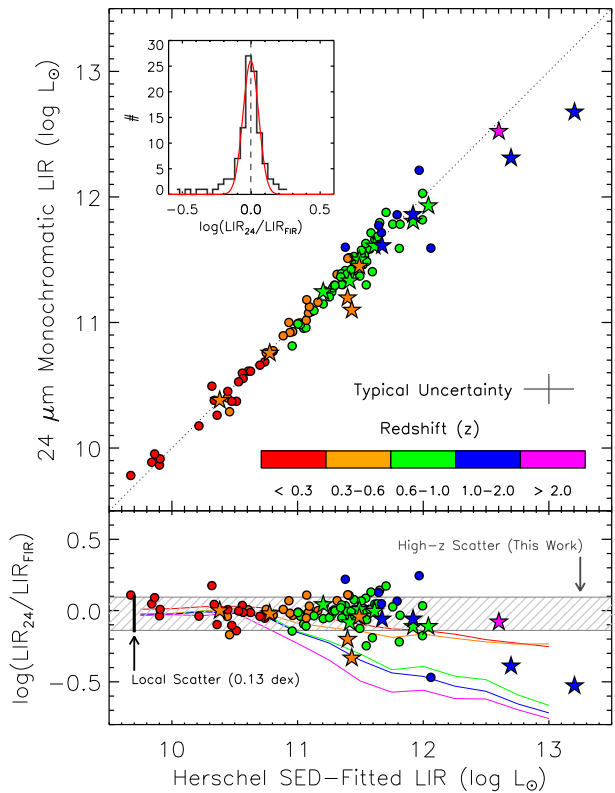}
\caption{Comparison of \LIR\ derived from the new 24 $\micron$
  indicator to \LIR\ measured by integrating the SED fitted to {\it
    Herschel} SPIRE ($250-500$ $\micron$) photometry for individual
  galaxies in the Extend {\it Chandra} Deep Field South (ECDFS) that
  have secure 
  spectroscopic redshifts (circles); and in the {\it Herschel} Lensing
  Survey (HLS; stars). The bottom panel shows the ratio of \LIR\ from
  our single-band 24 $\micron$ indicator to the far-IR \LIR. The
  overall scatter is 0.12 dex (shaded region), which is comparable to
  the scatter of the relationship between \LMIPS\ and \LIR\ found
  locally (0.13 dex; illustrated by a vertical bar on the left). The
  lines in the bottom panel (color-coded by $z$) indicate the ratios
  that would result if the star-forming regions in high-$z$ IR
  galaxies were as compact as those found locally (i.e., if our
  extended structure assumption fails), which is the case for 6\% of
  galaxies; each line corresponds to the upper-end of each $z$ bin in
  the top panel. The inset shows a Gaussian fit to the distribution,
  which has a $\sigma$ of 0.06 dex, indicating the degree of agreement
  where our extended structure assumption applies.}
\label{compare_lir_CDFS}
\end{figure}
\end{center}

We combine the {\it Spitzer} MIPS 24 and 70 $\micron$ catalog
from FIDEL and our SPIRE catalog to measure the \LIR\ for galaxies in
the ECDFS. We limit our sample to those that have high-quality
spectroscopic redshifts from the Arizona CDFS Environment Survey
(ACES) that has 5,080 secure redshifts \citep[$Q = 3$ or 4;
][]{Cooper12} to minimize the effects of 
redshift uncertainties. The
far-IR \LIR\ in the ECDFS is measured by fitting the \citet{Rieke09}
SED library to the MIPS 24 and 70 $\micron$ and at least one SPIRE
band. We redshift the SED library to the value determined by the
spectroscopic redshift of each object and minimize 
the $\chi^2$ value over the ranges of SED templates and the
normalization factors. We inspect each SED fit visually to ensure
fitting quality. Given the significantly smaller uncertainties of the 
MIPS photometry compared to those from SPIRE (especially at 24
$\micron$), we avoid the possibilities that too much weight could have
been put on the MIPS 24 and 70 $\micron$ points by adding 3\%
uncertainties in quadrature 
to each data point. We experimented with adding up to $25\%$ 
uncertainties in this way and found that the resulting \LIR\ values do
not depend sensitively on the choice of uncertainties added. In the
SED fitting, we require at least one data {\it Herschel} data point at
a rest-frame wavelength $> 30$ $\micron$ to be detected at $S/N >
5$. The final sample for SED fitting is selected at 250 $\micron$ flux
$> 7$ mJy, which yields 91 galaxies that passed all these criteria, in
addition to the blending avoidance criteria (Appendix A). The final
sample has redshifts ranging from $0.1 - 1.3$ with a median redshift
$z = 0.6$. The comparison between {\it Herschel} fiducial \LIR\ and
those from the new single-band 24 $\micron$ indicator is shown in
Figure \ref{compare_lir_CDFS}.

\subsubsection{{\it Hubble} Deep Field North Sample}
\begin{center}
\begin{figure}[ht]
\figurenum{4}
\epsscale{1.2}
\plotone{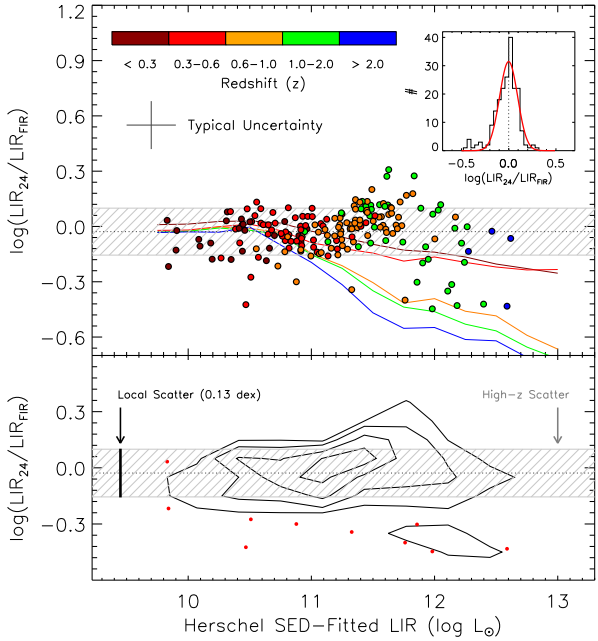}
\caption{Comparison of \LIR\ estimates from the new 24 $\micron$
  indicator to those measured with {\it Herschel} PACS and SPIRE
  ($100-500$ $\micron$) photometry in the {\it Hubble} Deep Field
  North. The top panel shows the ratio of \LIR\ from the 24 $\micron$
  indicator to the far-IR \LIR. The overall scatter is 0.13 dex
  (shaded region). Tracks representing ratios of compact objects, and
  the inset, are similar to those in Figure
  \ref{compare_lir_CDFS}. The bottom panel shows the contours
  encompassing the top 25$^{\rm th}$, 50$^{\rm th}$, 75$^{\rm th}$,
  90$^{\rm th}$ percentiles of the distribution, along with outliers
  in red. Comparison of results here and those in Figure
  \ref{compare_lir_CDFS} indicates that the addition of the shorter
  wavelength far-IR data (e.g., $100-160$) does not changed the result
  statistically, which demonstrates the general applicability of the
  indicator to far-IR-selected galaxies.\\}    
\label{compare_lir_HDFN}
\end{figure}
\end{center}

For the HDFN sample, we gathered all the data obtained by
{\it Herschel} in three different programs: the Guaranteed Time PACS
Evolutionary Probe (PEP, PI: D. Lutz), the {\it Herschel} Multi-tiered
Extragalactic Survey (HerMES, PI: S. Oliver), and the Open Time Key
Program GOODS-{\it Herschel} (PI: D. Elbaz). All the PACS and SPIRE
data taken by these surveys was downloaded from the {\it Herschel}
Science Archive (HSA) and merged together using HIPE and proprietary
dedicated software. The reduction steps were the standard for Level
2.0 and 2.5 data products in the HSA. Catalogs using MIPS 24~$\mu$m
priors and direct detections were built using the procedure described
in P\'erez-Gonz\'alez et al. (2010), which was conceived to extract
fluxes from faint sources that are difficult to detect directly and to
deblend nearby sources in the {\it Herschel} data (separated by more
than 
1 PSF FWHM for each band). The PACS catalogs reach 5$\sigma$
detections of 2~mJy at 100~$\mu$m, and 4~mJy at 160~$\mu$m. For SPIRE,
the 5$\sigma$ level is at 10, 14, and 17~mJy for the 250, 350, and
500~$\mu$m bands, respectively. We compiled spectroscopic redshift
measurements in the HDFN and augmented the sample with photometric
redshift measurements (P{\'e}rez-Gonz{\'a}lez et al. 2012, in
preparation) with accuracy of $\Delta z/(1+z) = 0.034$ based on a SED
fitting process described in \citet{Barro11}. Two hundred galaxies in
the HDFN are detected at $S/N > 5$ in at least one {\it Herschel} band
longward of rest-frame 30 $\micron$, passed the prior-based blending
avoidance criteria (Appendix A), and do not harbor X-ray or IR
power-law AGNs. Spectroscopic and photometric redshift measurements
are available for 137 and 63 of these galaxies, respectively; the
redshift range of the final sample is $0.06 - 2.21$, with a median
redshift $z = 0.7$. The \LIR\ comparison in HDFN is presented in
Figure \ref{compare_lir_HDFN}. 

\subsubsection{Gravitationally Lensed Galaxies Samples}
The HLS sample consists of 19 galaxies located behind the
Bullet Cluster at redshifts $0.4 < z < 3.24$ \citep{Rex10}. These
sources are detected in at least two {\it Herschel} bands (at $100 -
500$ $\micron$) and many are also observed in LABOCA 870 micron and
AzTEC 1.1 mm maps of the field \citep{Wilson08, Johansson10}. These
measurements tightly constrain the peak of the far-IR SED and
therefore provide accurate estimates of \LIR. We have excluded three
galaxies from the original HLS sample (Table 2 of Rex et al. 2010) in
our test: HLS12 and HLS13 ($z = 3.24$ and 2.9) because 24 $\micron$ no
longer traces PAH emission at their redshifts; HLS18, because its large
lensing magnification (54$\times$) is not well constrained due to
nearby objects \citep{Rex10}. Otherwise, the lensing magnifications in
the final HLS sample of 16 galaxies are small (median
1.1$\times$). These galaxies are shown as stars in Figure
\ref{compare_lir_CDFS}.

In addition to the HLS galaxies, we have tested the indicator on
individual 24 $\micron$-bright lensed star-forming galaxies at $1.0 <
z < 2.7$ for which we obtained near and mid-IR spectroscopic
observations with the Large Binocular Telescope and {\it Spitzer}, and
far-IR/sub-mm observations from the literature
\citep{Rujopakarn12}. The sample of five galaxies (Abell 2218b, Abell 
2218a, Abell 1835a, cB 58, and the Clone) is unique in that four
members are of LIRG luminosity. The gravitational lensing gives us
access to objects as low as $1.1 \times 10^{11}$ \Lsun\ (at a $z$ of
2.7). Although small, this sample thus provides an important
verification of the accuracy of our method for typical star-forming
galaxies at $z > 2$. \citet{Rujopakarn12} find that the 24 $\micron$
indicator from this work estimates \LIR\ in good agreement with their
far-IR \LIR\ values, with an average difference of 0.06 dex (although
there is one outlier whose difference is 0.18 dex). 

\begin{center}
\begin{figure}[ht]
\figurenum{5}
\epsscale{1.2}
\plotone{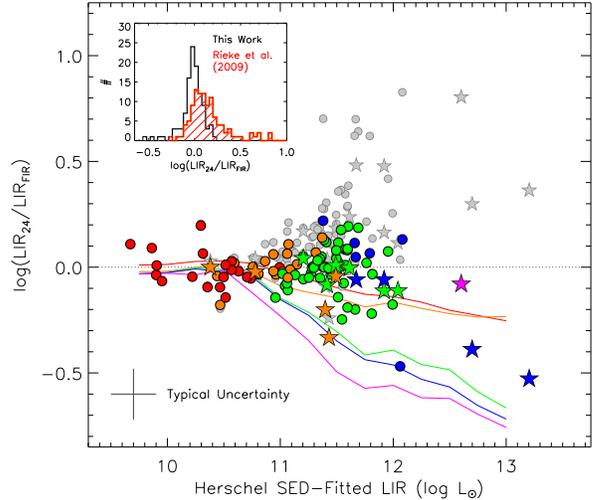}
\caption{Same as Figure \ref{compare_lir_CDFS} but with the \LIR\ values
  from the \citet{Rieke09} indicator shown as a comparison in
  grey. Color coding is identical to Figure \ref{compare_lir_CDFS} (color
  coding for the Rieke et al. indicator is omitted for clarity). The
  inset shows the histogram of the ratio, with values from the new
  indicator and those of \citet{Rieke09} shown in black and red
  histograms, respectively. While the \citet{Rieke09} indicator
  overestimates \LIR\ values compared to {\it Herschel} \LIR, this
  overestimation and the ``mid-IR excess problem'' (Section
  \ref{sec:indicator_testindiv}) are shown to be alleviated by the new
  \LIR\ indicator.\\}  
\label{compare_lir_rieke}
\end{figure}
\end{center}

\subsubsection{Summary of the \LIR\ Indicator Tests with Individual Galaxies}
We have found from the tests using the ECDFS, HDFN, and lensed
galaxies data that the systematic mid-IR excess issue discussed in the
Introduction is 
virtually removed. Figure \ref{compare_lir_rieke} uses the ECDFS
sample to illustrate the extent of the mid-IR excess when templates
for local galaxies are applied directly to high-$z$ galaxies, as is the
case in the formulae given by \citet{Rieke09}. The results from the
improved bolometric corrections are shown in comparison, which
indicate that the overestimation problem is no longer present. 

By using the new indicator, the resulting single-band 24
$\micron$-derived \LIR\ has an average agreement with the far-IR
observations of 0.02 dex in ECDFS (0.03 dex in HDFN) and a 
1-$\sigma$ scatter of 0.12 dex in the ECDFS, shown in Figure
\ref{compare_lir_CDFS} (0.13 dex in the HDFN, shown in Figure
\ref{compare_lir_HDFN}). The $0.12-0.13$ dex scatters are consistent
with the $0.13-$dex scatter of the relationship between \LMIPS\ and
\LIR\ in the local sample found by \citet{Rieke09}. The core
distribution of luminosity difference between \LIR\ estimates from
{\it Herschel} and the new indicator can be fitted by a Gaussian that
has a $\sigma$ of 0.06 and 0.09 dex in ECDFS and HDFN, respectively, 
reflecting the degree of agreement where the extended structure
assumption leads to a successful prediction of SED features. We found
8\% of the ECDFS and HLS galaxies (9\% of the HDFN galaxies) to have
\LIR\ estimates from this method disagreeing by more than 0.2 dex from
those of {\it Herschel}, with 2\% and 6\% on the over and
underestimation sides, respectively (2\% and 7\%, respectively, in the
HDFN). Upon inspecting the individual SED fits, we determined that the
overestimation is due to enhanced flux at 24 $\micron$ compared to
star-forming galaxy SEDs, suggesting a mid-IR emission contribution
from unidentified AGNs. SED fits for the 6\% of objects in ECDFS (7\%
in HDFN) where the indicator underestimates \LIR, 
indicates warmer far-IR SEDs than that predicted by the indicator,
which suggests that their starburst structures are more compact,
similar to local U/LIRGs, and inconsistent with our extended structure
assumption. This is illustrated by lines (color-coded by $z$) in
Figures \ref{compare_lir_CDFS}, \ref{compare_lir_HDFN}, and
\ref{compare_lir_rieke} that show where the compact U/LIRGs would lie
if they are assumed (incorrectly) to have extended structure. Figures
\ref{compare_lir_CDFS} and \ref{compare_lir_HDFN} indicate that,
although some U/LIRGs at high $z$ could be compact, these are in the
minority at a $10\%$ level, which suggests that our assumption that
high-$z$ star-forming galaxies indeed have extended star formation
(i.e., that they are dominated by main-sequence star formation) is
consistent with the properties of individual galaxies tested here.

We note that the fraction of compact objects here should be taken as
an upper limit because far-IR selection of galaxies, such as
presented, especially at $z > 2$, could be biased toward objects with
warmer thermal dust emission and may cause compact objects to be
overrepresented.

\begin{center}
\begin{deluxetable*}{ccccccccccc}
\tablewidth{0pt}
\tablecaption{The Luminosity of the Appropriate \citet{Rieke09} SED
  Template to Describe SED of a Star-Forming Galaxy for a Given 24
  $\micron$ Flux and Redshift}
\tablehead{\colhead{f$_{24}$} & \multicolumn{10}{c}{Redshift ($z$)} \\
\colhead{(mJy)} & \colhead{0.10} & \colhead{0.50} & \colhead{0.75}
& \colhead{1.00} & \colhead{1.25} & \colhead{1.50} & \colhead{1.75} &
\colhead{2.00} & \colhead{2.50} & \colhead{2.80}} 
\startdata
0.02 & $\ldots$ &   9.9 &  10.1 &  10.2 &  10.4 &  10.5 &  10.5 &  10.5 &  10.6 &  10.8 \\
0.05 & $\ldots$ &  10.1 &  10.3 &  10.4 &  10.6 &  10.7 &  10.7 &  10.7 &  10.8 &  11.0 \\ 
0.1 & $\ldots$ &  10.3 &  10.5 &  10.6 &  10.8 &  10.9 &  10.8 &  10.8 &  11.0 &  11.1 \\ 
0.2 & $\ldots$ &  10.4 &  10.6 &  10.7 &  10.9 &  11.0 &  11.0 &  11.0 &  11.1 &  11.2 \\ 
0.4 & 9.7 &  10.6 &  10.8 &  10.9 &  11.1 &  11.1 &  11.1 &  11.1 &  11.3 &  11.4 \\ 
0.8 & 9.9 &  10.7 &  10.9 &  11.0 &  11.2 &  11.3 &  11.2 &  11.2 &  11.4 &  11.5 \\ 
1.5 &10.0 &  10.9 &  11.0 &  11.1 &  11.3 &  11.4 &  11.3 &  11.3 &  11.5 &  11.6 \\ 
3.0 & 10.2 &  11.0 &  11.1 &  11.3 &  11.4 &  11.5 &  11.4 &  11.5 &  11.6 &  11.7
\enddata
\tablecomments{Dots indicate that the combination of flux and
  redshift would yield a SED template that is outside the luminosity
  range of the \citet{Rieke09} SED library, \LIR\ $ = 10^{9.75} -
  10^{13}$ \Lsun. As a reference, the log(\LIR/\Lsun) values for the
  archetypal local IR galaxies M82 and Arp 220 are 10.77 and 12.21,
  respectively.}
\label{table_fluxztemplate}
\end{deluxetable*}
\end{center}

\begin{figure*}
\figurenum{6}
\epsscale{1.15}
\plottwo{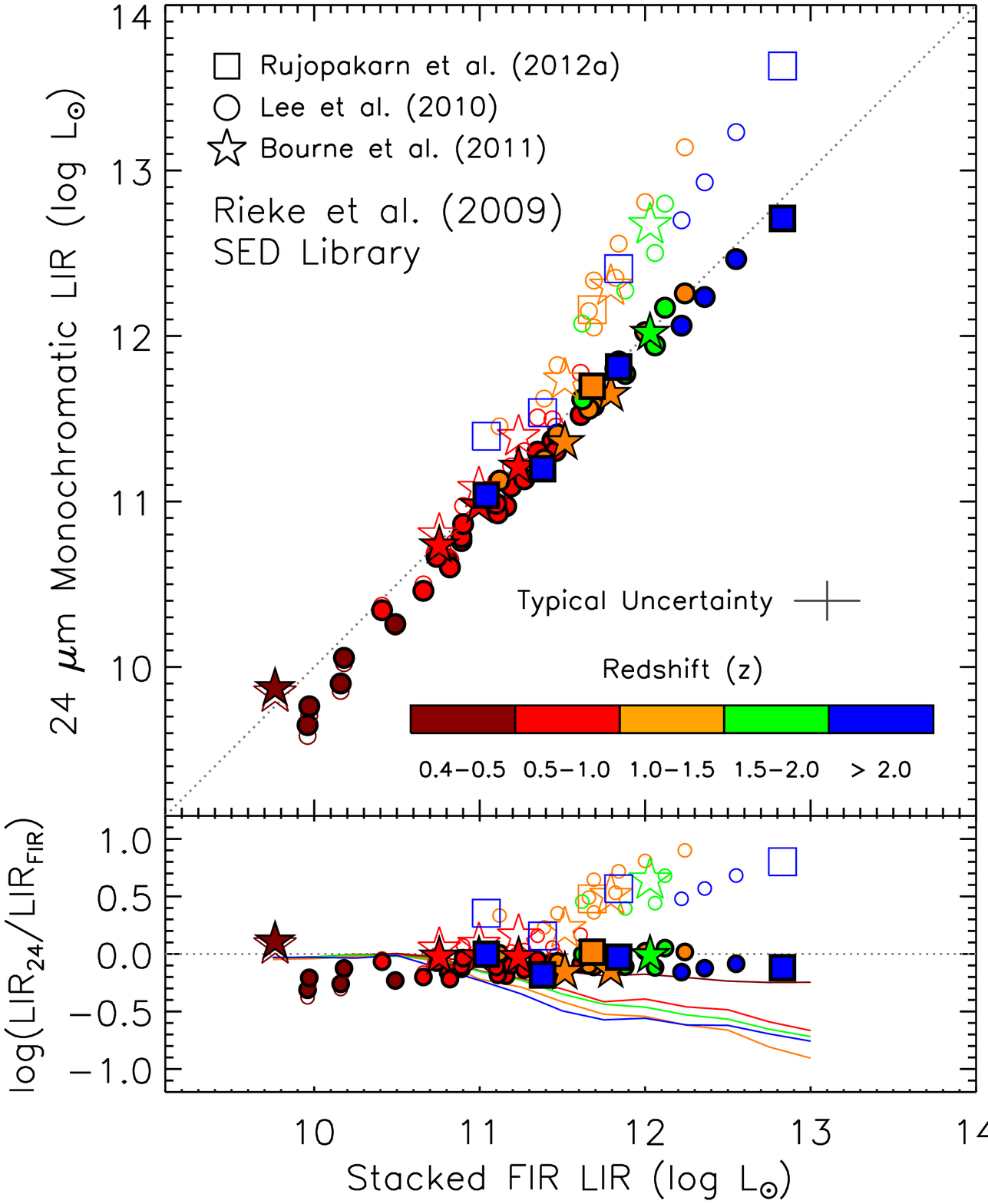}{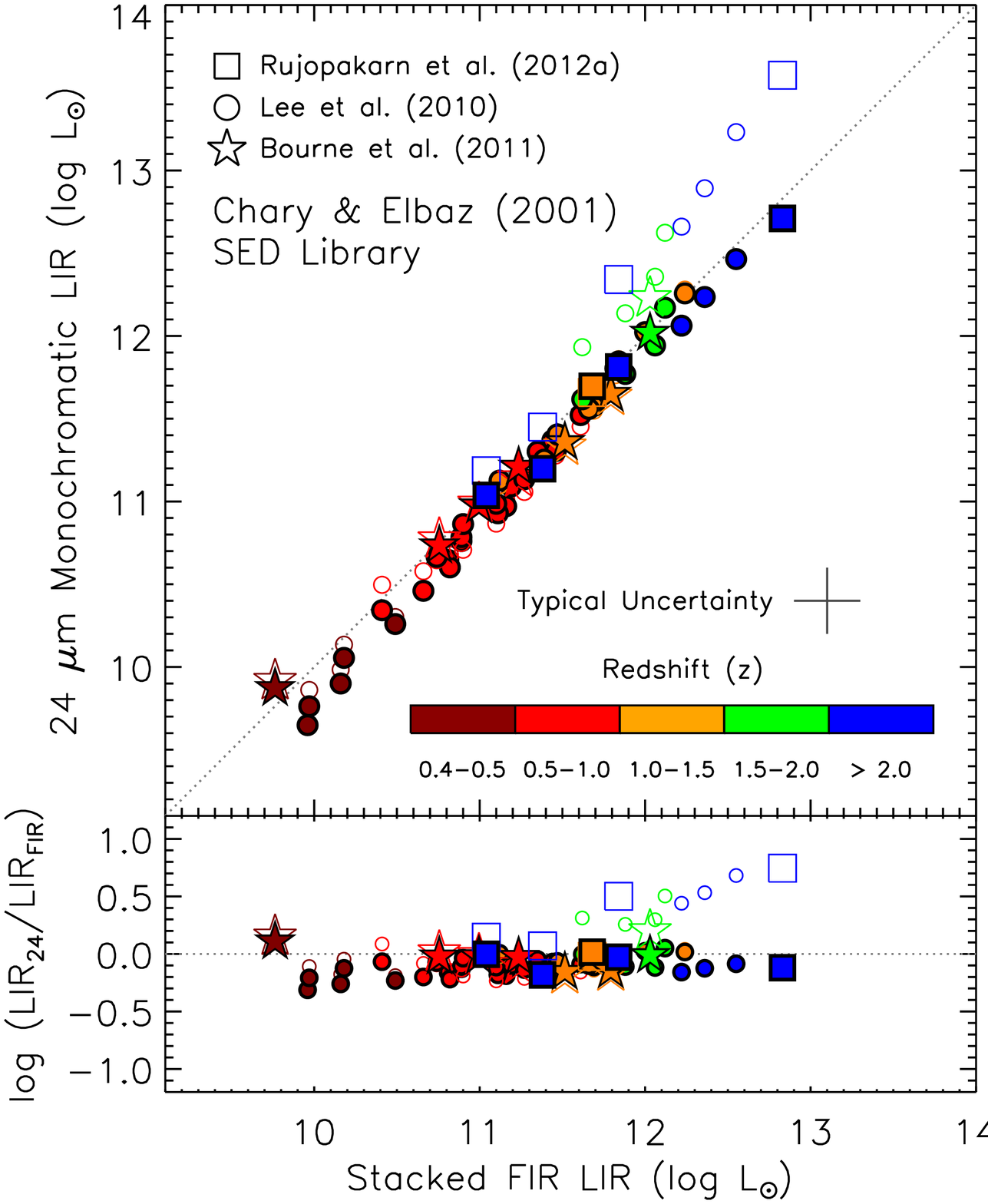}
\caption{Comparison of \LIR\ from our single-band 24 $\micron$
  indicator to \LIR\ measured from stacked and gravitationally lensed
  galaxies with far-IR observations. The comparison samples are from
  \citet{Lee10} who determined \LIR\ at $0 < z < 3$ by stacking 70
  $\micron$ and 160 $\micron$ observations of over 35,000 COSMOS
  galaxies (selected at 24 $\micron$); from  \citet{Bourne11} whose
  \LIR\ was determined from stacking a broad range of observation from
  24 $\micron$ to 610 MHz of 3,172 galaxies at $0 < z < 2$ in the
  ECDFS-FIDEL sample (selected at 3.6 and 4.5 $\micron$); and
  from  \citet{Rujopakarn12} whose sample consists of gravitationally
  lensed galaxies at $1.0 < z < 2.7$. In the left and right
  panels, we show  \LIR\ estimates from the new indicator as filled
  symbols in comparison to \LIR\ estimates from the R09 and CE01 SED
  library (open symbols in each panel). The lines in the bottom plot
  of the left panel illustrate the ratios that would result if the
  star-forming regions in high-$z$ IR galaxies were as compact as those
  found locally (see also, Figures \ref{compare_lir_CDFS}). The right
  panel also illustrates our findings discussed in  
  Section \ref{sec:indicator_teststat} that the mid-IR excess problem
  does not become apparent for the CE01 estimates until $z \sim 
  1.5$ because their overestimation of the PAH-region fluxes compared
  to local galaxies helps compensate the evolution of PAH
  strength at high $z$.}
\label{compare_lir_leebourne}
\end{figure*}

\subsection{Tests of the 24 $\micron$  \LIR\ Indicator on Average
  SEDs}\label{sec:indicator_teststat}  
Next, we investigate the applicability of the new indicator on large
galaxy survey samples using stacked photometry to represent the average
properties of galaxies. We employ two samples for these tests: (1) the
stacked 70 $\micron$ and 160 $\micron$ photometry of a 24
$\micron$-selected sample  from \citet{Lee10} in COSMOS; and (2) the
stacked 24 $\micron$, 70 $\micron$, 160 $\micron$, 1.4 GHz, and 610
MHz observations of a NIR-selected sample from the ECDFS
\citep{Bourne11}. Unlike the earlier ECDFS, HDFN, and HLS samples,
these \LIR\ estimates rely to a greater extent on extrapolating a
fitted-SED to the dust emission peak. The goal of this section is thus
not to determine the scatter for individual galaxies, but to test the
indicator on a large sample across broad ranges of $z$ and \LIR.

The \citet{Lee10} sample stacks 70 $\micron$ and 160 $\micron$
observations of galaxies selected at 24 $\micron$ in COSMOS and
represents the average SED properties of over 35000 galaxies. The
stacks were done in bins of 24 $\micron$ flux ($0.06 < f_{24} < 3.00$
mJy) and redshift ($0 < z < 3$). Lee et al. then fit SEDs to the
average 24 $\micron$, 70 $\micron$ and 160 $\micron$ fluxes in each 
stacked bin to estimate \LIR. We
have excluded bins where the fraction of X-ray sources (i.e., AGN)
exceeds 10\% based on their Figure 3. For the actual \LIR\ calculation
we use the average value of 24 $\micron$ flux and 
redshift in each bin (N. Lee, 2011, private communication) to
calculate the average \LIR\ of the bin, which can be compared with the
estimates from \citet{Lee10}.

The \citet{Bourne11} stacked sample of 3172 galaxies was selected
using {\it Spitzer} IRAC 3.6 $\micron$ and 4.5 $\micron$ photometry
(i.e., a stellar mass selected sample) in the ECDFS. The selection in the
near-IR provides an independent sample that has a 
potential to reveal selection effects (if any) inherent to mid- and
far-IR selections (e.g., the \citet{Lee10} and \citet{Kartal10}
samples). Bourne et al. stacked observations of the sample in 
7 redshift bins from $0 < z < 2.0$. These observations include 24
$\micron$, 70 $\micron$, 160 $\micron$, 1.4 GHz and 610 MHz; the
stacked fluxes of the {\it Spitzer} MIPS bands were then used to fit a
M51 SED to estimate \LIR.

We compare with \citet{Lee10} and \citet{Bourne11} for our indicator
and for the R09 and CE01 SED libraries separately in the left and
right panels of Figure \ref{compare_lir_leebourne}, respectively. A
difference between the two panels is that the mid-IR excess becomes a
problem for the R09 library well below $z = 1$ while it only becomes
noticeable for CE01 templates at $z > 1.5$ (see also, Elbaz et
al. 2010, Figure 1), as is observed in Figure
\ref{compare_lir_CDFS}. This is because the omission of silicate
absorption features in the CE01 SED templates results in an
overestimation of PAH-region flux in the local SED templates that
coincidentally helps compensate for the evolution of PAH strength and
delays the emergence of the mid-IR excess problem until beyond $z \sim
1.5$. The R09 templates correctly include the silicate absorption
locally, and hence suffer a larger overestimation of L(TIR). We have
now shown that the R09 templates can be used to describe the SEDs of
galaxies and have estimated the bolometric corrections out to $z \sim
2.8$ given an appropriate choice of SED template based on the \LIRSD.

\subsection{Verifying the Uncertainty Estimate for Individual Galaxies
  with an Independent Sample}\label{sec:indicator_testcosmos}
Having established in Section \ref{sec:indicator_testindiv} that the
scatter of the \LIR\ estimates from our indicator is roughly $0.12 -
0.13$ dex compared to far-IR \LIR\ measurements based on the ECDFS,
HDFN, and HLS data, we now seek to verify this scatter estimate using
a larger sample selected at a different wavelength. 

We compare the new \LIR\ indicator to the \LIR\ estimates from the 70
$\micron$-selected COSMOS sample \citep{Kartal10}. The entire sample
contains 1503 galaxies at $0 < z < 3.5$ (median $z = 0.5$) with
\LIR\ estimated by fitting IR photometry from {\it Spitzer} at the 8,
24, 70, and 160 $\micron$ bands to a collection of SED libraries, with
the best fit selected via $\chi^2$ minimization among the
\citet{CharyElbaz01}, \citet{DaleHelou02}, \citet{Lagache03}, and
\citet{Sieben07} templates. For our comparison, we exclude sources
with X-ray luminosity $> 10^{42}$ erg/s and those with
radio-excess. The sample is complete at 70 $\micron$, by definition,
and also at 24 $\micron$, which allows the 70 and 24 $\micron$ color
to be used as a criterion to select AGN that dominate the mid-IR
emission. Specifically, we exclude objects with log(f$_{70}$/f$_{24}$)
$<$ $0.2z +0.7$ (i.e., objects with enhanced 24 $\micron$ flux for a
given 70 $\micron$ flux compared to the SEDs of star-forming
galaxies), which removes 86 galaxies. Above $10^{12.5}$ \Lsun, the
\citet{Kartal10} sample contains very few sources that do not harbor
AGN or QSO. We limit our comparison sample to those with uncertainties
in SED-fitted \LIR\ $<0.35$ dex to avoid comparing to objects with
uncertain luminosity.

Among the 1503 galaxies selected at 70 $\micron$, 463 are
detected at 160 $\micron$ and their fluxes have been included
in the SED fitting by \citet{Kartal10}; 410 of
these pass our AGN and uncertainty criteria. However,
the 160 $\micron$ detections for these sources are of low significance:
236/410 galaxies (57\%) are below 5 $\sigma$ where
the $\sigma$ value includes the confusion noise of 10 mJy for
the survey \citep{Frayer09}. Although a consistent \LIR\ is
obtained if a stacked 160 $\micron$ flux is used for SED fitting,
including these fluxes in the individual fits results in values typically
high by 0.2 dex \citep{Kartal10}. This problem is consistent with
\citet{Hogg98}, who show that detections below $4-5 \sigma$ are biased
towards brighter fluxes than their true values; see their Figure 2;
this effect is commonly known as the Eddington bias. We therefore
exclude the 160 $\micron$-detected objects.

After applying the cuts discussed at the beginning of this section, we
have 751 sources left. These sources have redshifts ranging from $z =
0.07$ to $1.81$ with a mean and median redshift of 0.52 and 0.43. The
luminosity of this subsample ranges from \LIR\ of $10^{9.5}$ to
$10^{12.5}$ \Lsun\ and the mean and median \LIR s are $10^{11.1}$ and
$10^{11.2}$ \Lsun, respectively. 

The comparison with \LIR\ estimated from the 24 $\micron$ fluxes alone
indicates an average scatter of $0.25$ dex relative to the assigned
\LIR\ values. Some fraction of this scatter must arise in the assignment of
\LIR\ by \citet{Kartal10}; the median luminosity uncertainty within
their work is 0.23 dex. That is, 
the 24 $\micron$-only calculation agrees with their multi-wavelength 
fits virtually within the internal scatter of these fits. We quantify
the contribution of uncertainties intrinsic to the new indicator in
this test by conducting a Monte Carlo experiment to determine the
scatter that must arise from our estimation of \citet{Kartal10}
luminosities. We simulate a sample ($n = 10^4$) with a scatter of 0.23
dex and re-measure the values, introducing measurement errors in the
process, which shows that an uncertainty of $\sim$0.1 dex associated
with the new indicator will broaden the intrinsic scatter of the
\citet{Kartal10} sample to the 0.25 dex measured. This result is
consistent with the $0.12-0.13$ dex scatter found in Section
\ref{sec:indicator_testindiv}.

\subsection{Summary of the Tests}\label{sec:indicator_testsummary} 
We tested the new 24 $\micron$ indicator on far-IR \LIR\ measurements
of seven samples of galaxies selected with various techniques. These
include (1) 91 galaxies in ECDFS selected at 
$250-500$ $\micron$; (2) 200 galaxies in HDFN selected at $100-500$
$\micron$; (3) 751 galaxies in COSMOS selected at 70 $\micron$
\citep{Kartal10}; (4) 16 far-IR-bright galaxies from
\citet{Rex10}; (5) five 24 $\micron$-bright lensed galaxies from
\citet{Rujopakarn12}; (6) stacked photometry of 35,000 galaxies in
COSMOS selected at 24 $\micron$ from \citep{Lee10}; (7) stacked
photometry of 3172 galaxies in the ECDFS selected at 3.6 and 4.5
$\micron$.  

All of these tests indicate that the new indicator has eliminated the
systematic overestimation of \LIR\ due to mismatching of SED
templates, which is caused by the misassignment of the SEDs of compact
local U/LIRGs to the extended U/LIRGs at high $z$ at the same
\LIR. For star-forming galaxies at $0.0 < z < 2.8$, the new indicator
yields \LIR\ estimates consistent with {\it Herschel} far-IR
measurements with an average agreement of $0.02-0.03$ dex and a
$1-\sigma$ scatter of $0.12-0.13$ dex. Based on the samples in ECDFS
and HDFN, we estimate that the fraction of compact merger-triggered
U/LIRGs beyond the local Universe to be $\sim$10\% (more
discussion in Section \ref{sec:discuss_sfmodes}).

We tabulate the luminosities of the recommended R09 SED templates to 
describe star-forming galaxies given their observed 24 $\micron$
fluxes and redshifts in Table \ref{table_fluxztemplate}. Even at the
bright-end of the flux range at high-$z$ (e.g., $f_{24} = 3.0$ mJy at
$z = 2.8$, which corresponds to \LIR\ of $2 \times 10^{14}$ \Lsun),
the appropriate R09 templates are those of local LIRGs with \LIR\ of
no more than $5 \times 10^{11}$ \Lsun. In fact, it is evident from the
table that most IR-luminous star-forming galaxies at $0 < z < 2.8$
exhibit spectral characteristics of local galaxies with \LIR\ in the
range of $10^{10} - 3 \times 10^{11}$ \Lsun.\\

%
%
%
%

\section{\LIR$-$SFR RELATION}\label{sec:final_SFR}
Finally, SFRs can be determined by making use of the relationship
between \LIR\ and the rest-frame \LMIPS, and subsequently the
rest-frame \LMIPS\ and SFR, originally given by \citet{Rieke09}. The introduction
of the stretching factor, which effectively re-normalizes the SED
templates, requires a modification of the relationship between
\LIR\ and \LMIPS. The original fit as given in equation A6 of
\citet{Rieke09} is 
\begin{equation}
\label{eq6}
\log L({\rm TIR})_{z = 0} = 1.445 + 0.945 \log L(24~\mu {\rm m},\;L_\odot)
\end{equation}

\noindent The modified relationship is obtained by re-fitting
equation \ref{eq6} with the stretching factor, $S_i$, multiplying both
\LIR\ and \LMIPS\ for each template $i$. The re-fitted relation allows \LMIPS\ to be calculated by substituting $L({\rm TIR})_{{\rm new}}$ (from equation \ref{eq5}) in the following.
\begin{equation}
\label{eq7}
\log L(24~\mu {\rm m},~L_\odot) = \frac{1}{0.982} \left[ \log L({\rm TIR})_{{\rm new}} - 1.096 \right]
\end{equation}

To determine the SFR from the rest-frame \LMIPS, the calibration given by
\citet{Rieke09} remains valid. However, that calibration has a term
that corrects for a decrease in \LMIPS/\LIR\ ratio above \LIR\ $= 10^{11}$
\Lsun. Since the correction is motivated by an increase of the optical
depth at high 
\LIR\ that prevents the mid-IR emission from escaping, the threshold
at which optical depth becomes significant depends directly on the
geometry of the galaxy. In the same way that the extended structure of
the galaxy beyond the local Universe affects the IR-emitting
environment, the optical depth will consequently be lower for a given
\LIR\ and the luminosity threshold where the optical depth should
apply has to be scaled up by a stretching factor as well. The $S_i$
corresponding to the original threshold is 12.6$\times$ (referring to
Table \ref{table_si}), yielding an IR luminosity threshold of $1.3 \times 
10^{12}$ \Lsun, which is equivalent to \LMIPS\ of $1.6 \times 10^{11}$
\Lsun. Therefore the relationship between SFR and \LMIPS\ (from Equation \ref{eq7}) is given by 
\begin{eqnarray}
{\rm SFR}({\rm M}_\odot \;{\rm yr}^{-1}) =7.8\times 10^{-10}\;{\rm
  L}(24~\mu {\rm m},\;L_\odot ) \nonumber 
\end{eqnarray}
\noindent
for $5 \times 10^9$ \Lsun\ $\le$ \LIR\ $\le 1.3 \times
10^{12}$ \Lsun\ or $6 \times 10^8$ \Lsun\ $\le$ \LMIPS\ $\le
1.6 \times 10^{11}$ \Lsun. For \LIR\ $> 1.3 \times 10^{12}$ 
\Lsun\ or \LMIPS\ $> 1.6 \times 10^{11}$ \Lsun, 
\begin{eqnarray}
\label{eq8}
{\rm SFR}({\rm M}_\odot \;{\rm yr}^{-1})=7.8\times 10^{-10}\;{\rm
  L}(24~\mu {\rm m},\;L_\odot ) \nonumber \\  \times \left[6.2\times 10^{-12}\;{\rm
    L}(24~\mu {\rm m},\;L_\odot)\right]^{0.048} 
\end{eqnarray}

This calibration is based on that of \citet{Kennicutt98} but with the
\citet{Kroupa02} IMF, which yields SFRs a factor of $0.66$ 
of those assuming the \citet{Salpeter55} IMF. \\

To summarize, the recipe to estimate \LIR\ and SFR from a given set of
24 $\micron$ flux and redshift measurements is following. An IDL
implementation of the steps below is available at our
website\footnote[1]{http://ircamera.as.arizona.edu/rujopakarn2013}. 
\smallskip

(1) Check that the galaxy of interest does not harbor luminous 
AGN that dominates its 24 $\micron$ emission using the mid-IR
\citep[e.g.,][]{Donley12} or X-ray criteria (e.g., by requiring L$_{\rm
  X}[0.5-8.0~{\rm keV}] < 10^{42}$ erg/s).
\smallskip

(2) To calculate total IR luminosity, interpolate for the coefficients
$A'(z)$ and $B'(z)$ from Table \ref{table_azbz} and calculate the
luminosity distance for the object's redshift, then substitute
these values in Equation \ref{eq5} along with the observed 24
$\micron$ flux in mJy. The resulting $L({\rm TIR})_{{\rm  new}}$ is
in unit of \Lsun. 
\smallskip

(3) To calculate SFR (in \Msun/yr), first calculate the rest-frame
\LMIPS\ in \Lsun\ using Equation  \ref{eq7}, then substitute it in
Equation \ref{eq8} at the appropriate luminosity range. \\

%
%
%
%

\section{DISCUSSION}\label{sec:discuss}
In Section \ref{sec:discuss_validity}, we discuss the validity of
using 24 $\micron$ observation at $0 < z < 2.8$ to estimate \LIR\ and
SFR; in Section \ref{sec:discuss_sfmodes}, we discuss the implications
of the \LIR\ indicator in light of the test results from Section
\ref{sec:indicator_test}. In addition, we will discuss the
implications of the new indicator on the maximum typical \LIR\ for
high-$z$ ULIRGs in Section \ref{sec:discuss_LEdd}.

\subsection{Validity of the Aromatic Emission as a
  \LIR\ and SFR Indicator}\label{sec:discuss_validity}   
The success of the new indicator, which  utilizes the luminosity at
rest-frame wavelengths of 24 $\micron$ to 6 $\micron$ at $z = 0$ to
2.8, indicates that photometry dominated by aromatic emission is a
surprisingly good tracer for \LIR. We discuss this result in terms of
two major factors affecting the aromatic feature strength: the
metallicity dependence and the presence of AGN.

\begin{figure*}
\figurenum{7}
\epsscale{1.125}
\plottwo{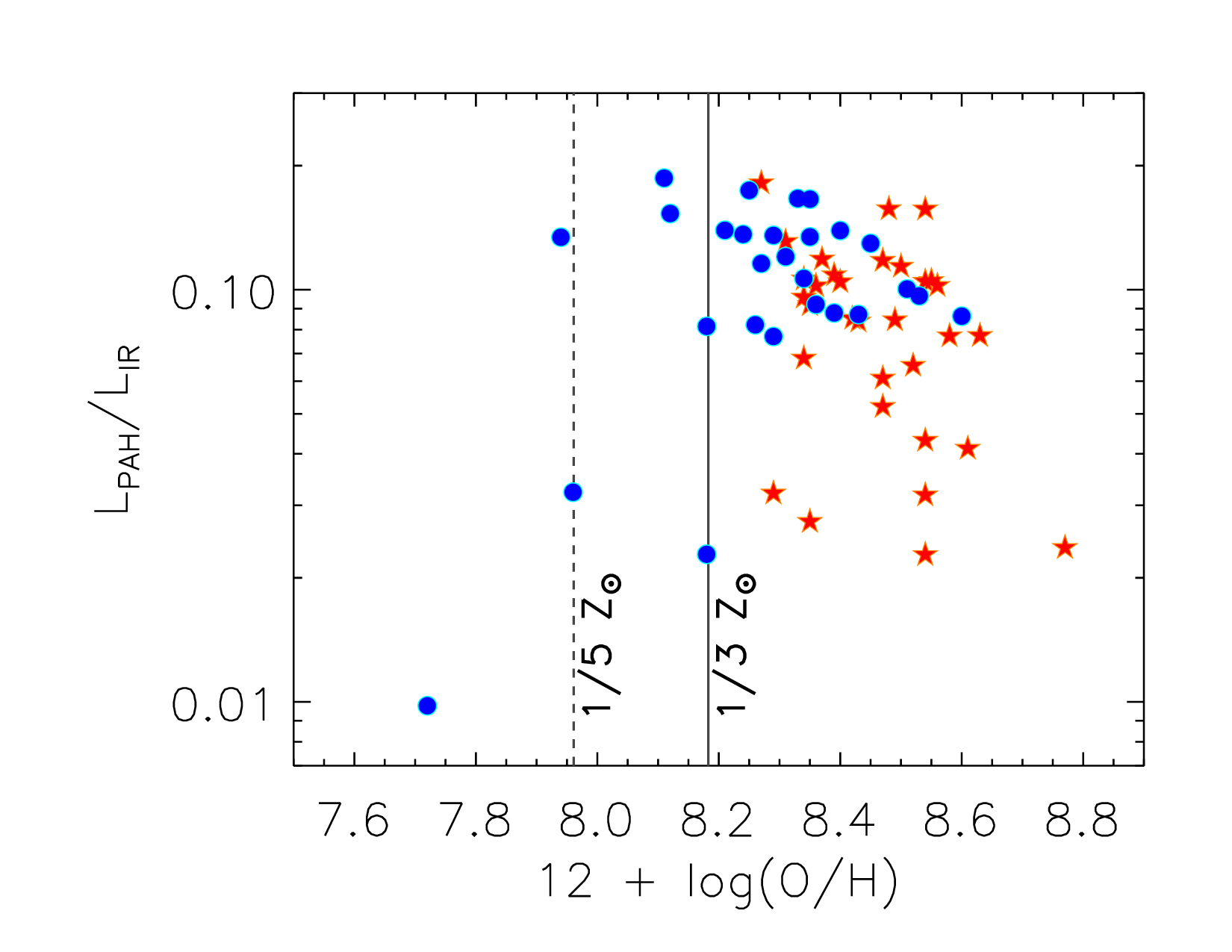}{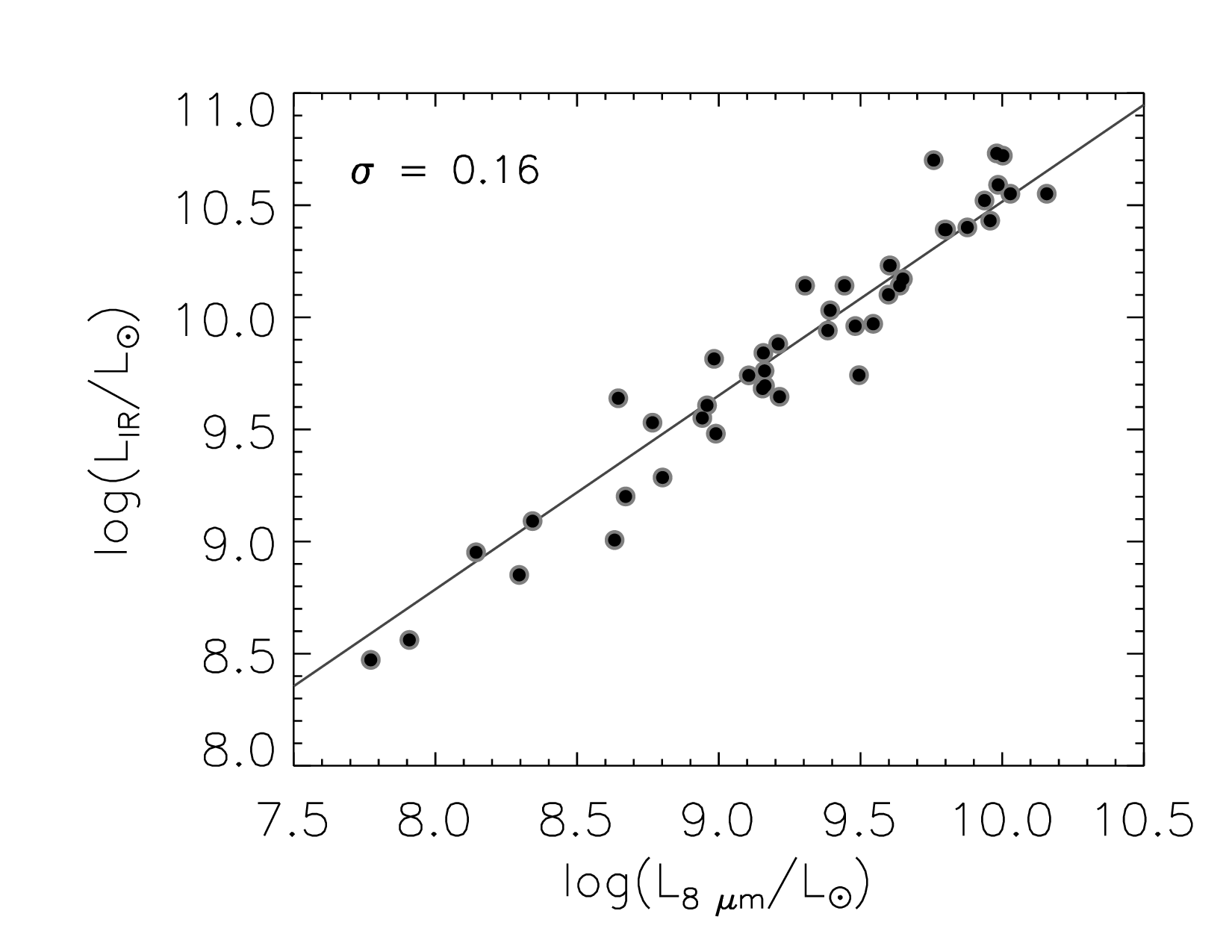}
\caption{(Left) The sources of scatter in using aromatic
  emissions (e.g., PAH) to predict \LIR\ and SFR are primarily the 
  metallicity dependence and the presence of AGN. Blue dots and red
  stars show the ratios of aromatic luminosity to \LIR\ as a function
  of metallicity for HII and AGN dominated local galaxies,
  respectively \citep{JDSmith07}. For HII-dominated galaxies with
  metallicity greater than $1/3 ~Z_{\odot}$ (solid vertical line),
  which is similar to the environment of high-$z$ star-forming
  galaxies, the correlation of aromatic luminosity, L$_{{\rm PAH}}$, and
  \LIR\ has a $\sim0.1$ dex scatter, suggesting that aromatic emission
  could serve as a good \LIR\ and SFR indicators at high
  $z$. (Right) The correlation between the rest-frame
  L$(8~\micron)$ and \LIR\ for local star-forming galaxies that are
  dominated by HII-emission and have high metallicity, similar to those
  expected at high $z$, has a scatter of $0.14$ dex. The small
  scatter underlines the potential of using aromatic emission as a
  \LIR\ and SFR indicator.}
\label{eightmicron_lir}
\end{figure*}

The correlation of the aromatic luminosity to \LIR\ was studied by
\citet{Rigopoulou99}, \citet{Roussel01} and \citet{Elbaz02} using {\it
  ISO}; later by \citet{Wu05} with {\it Spitzer}; and recently by
\citet{Elbaz11} with {\it Herschel}. Although large scatter in L(8
$\micron$)/\LIR\ is apparent over a large range of metallicity
\citep{Calzetti07}, for high-metallicity systems ($Z > 1/3$
$Z_{\odot}$, which is equivalent to $12 + $log(O$/$H) $\gtrsim 8.2$),
\citet{Calzetti10b} reports that the stellar-continuum-subtracted PAH
emission shows a good correlation with the SFR. Furthermore,
\citet{Engelbracht08} and \citet{JDSmith07} find that the ratio of
aromatic luminosity to \LIR\ does not vary significantly at
metallicity $\gtrsim 1/3~Z_{\odot}$. Galaxies at $z \sim 2$ have on
average $0.3$ dex lower metallicity than local galaxies
\citep{Erb06}. Therefore, galaxies more massive than $3 \times 10^{9}$
M$_{\odot}$ should be sufficiently metal rich ($> 1/3~Z_\odot$) that
the \LIR$/$PAH calibration is valid. If we consider the ratio of
aromatic-luminosity-to-\LIR\ as a function of metallicity in the left
panel of Figure \ref{eightmicron_lir} within the range of metallicity
expected at high-$z$, the scatter of the relationship between aromatic
luminosity and \LIR\ is in fact about $0.1$ dex in the
non-Seyfert/LINER sample. This suggests aromatic luminosity to be a
good indicator for \LIR\ at high-$z$ given that an effort is made to
exclude AGN from the sample (such as the AGN exclusion criteria
employed in tests in Sections \ref{sec:indicator_testindiv} and
\ref{sec:indicator_teststat}). 

\citet{JDSmith07} found that AGN can
significantly suppress the PAH emission \citep[see
  also,][]{Moorwood86, Roche91, Genzel98}.  The suppression 
of PAH luminosity is shown in Figure \ref{eightmicron_lir} (left)
using the total PAH luminosities measured by \citet{JDSmith07}
and a metallicity measurement from \citet{Moustakas10}. Moderate
luminosity AGN, however,
do not affect the PAH emission of the entire host
galaxy. \citet{DiamondStanic10} use {\it Spitzer} IRS to compare
nuclear and non-nuclear spectra of nearby Seyfert galaxies and found
that while the PAH features in the nuclear spectra are clearly
suppressed, the features are of normal strength in the
outer disk. Therefore, with regard to using PAH emission to trace
\LIR\ in galaxies hosting moderate luminosity AGN, the PAH
suppression due to AGN is likely limited 
to the nuclear region and PAH emission arising from the rest of
the galaxy should still provide a good tracer of the SFR.

We further test the validity of aromatic emission (e.g., 8 $\micron$
rest-frame observation) as a measure of \LIR\ in a sample of
HII-dominated galaxies with moderate to high metallicity in the right
panel of Figure \ref{eightmicron_lir}. In this Figure, the 8 $\micron$
observations are from {\it Spitzer} and the \LIR\ estimates are based
on the {\it IRAS} all-sky survey using 12, 25, 60, and 100 $\micron$
observations \citep{Sanders03}. This sample is chosen such that it
mimics the population of star-forming galaxies expected at high
$z$: high-metallicity normal star-forming galaxies without
AGN. The scatter of the relationship between L(${8~\micron}$) and
\LIR\ in Figure \ref{eightmicron_lir} is $0.14$ dex with the ratio
\LIR/L(${8~\micron}$) of $4.3 \pm 1.6$. \citet{Elbaz11} reports a
similar \LIR/L(${8~\micron}$) of $4.9^{+2.9}_{-2.2}$ for their
GOODS-{\it Herschel} sample out to $z \sim 3$ (that is, a 1-$\sigma$
spread of $\sim$ 0.23 dex) without applying corrections for the shape
of the SED \citep[see also,][]{Nordon11}.

\subsection{Modes of Star Formation and the Validity of the Main
  Sequence 
  Assumption}\label{sec:discuss_sfmodes}
In the process of constructing our \LIR\ indicator, we have assumed
that all non-local star-forming galaxies lie in the main-sequence,
i.e., have extended star formation. In other words, that the star
formation beyond the local Universe is not dominated by compact,
merger-induced, nuclear-concentrated starbursts. The results from
Sections \ref{sec:indicator_testindiv} and
\ref{sec:indicator_teststat} indicate that our assumption is
consistent with the nature of high-$z$ galaxies: (1)
the potentially compact starbursts, whose \LIR\ are underestimated
compared to their far-IR \LIR, comprise only $6-7\%$ of the sample,
and (2) the comparison of stacked samples shows no systematic
discrepancies.

At low redshifts ($z \lesssim 0.5$), the bottom panel of Figure
\ref{compare_lir_CDFS} and the top panel of Figure
\ref{compare_lir_HDFN} suggest that the \LIR\ values from the new
indicator and from that of \citet{Rieke09}, which assumes
compact-starburst (i.e., local) SED templates without the use of
\LIRSD, converge to within 0.25 dex of each other at all \LIR\ up to
$10^{13}$ \Lsun. The agreement is better at lower \LIR, e.g., $\sim
0.1$ dex at $10^{12}$ \Lsun\ that is typical for star-forming U/LIRGs
at low $z$. That is, for practical purposes, both indicators can be
used interchangeably below $z \sim 0.5$ regardless of the compactness of
galaxies' star-forming regions.

At intermediate $z$ ($0.5 \lesssim z \lesssim 1$), the precise
fractions of galaxies that form stars in the extended and compact
modes remain a subject of controversy. \citet{Lotz11} discuss the
various merger indicators critically to remove the discrepancies as
much as possible and find that mergers are, in fact, important out to
$z \sim 1.5$, with minor mergers about three times more frequent than
major ones. To the first order, this result appears consistent with
the tests of our \LIR\ indicator in Section
\ref{sec:indicator_teststat}, which support our initial assumption
that a majority of star-forming galaxies beyond the local Universe 
reside in the main sequence of star-forming galaxies. The main
sequence picture for IR galaxies is also advocated by \citet{Elbaz11}
who find that nearly $80\%$ of IR star-forming galaxies studied by
{\it Herschel} have the ratio of the 8 $\micron$ luminosity to
\LIR\ consistent with being in the extended mode of star formation. A
remaining issue is the precise transition scheme from the
merger-induced mode of star formation found locally to the extended
mode of star formation at high $z$, which is actually among the
fundamental questions in galaxy evolution as it is closely tied to the
mechanism of the decline of SFR since $z \sim 1.5$. A theoretical
interpretation was provided by \citet{Hopkins10}, but future
observations will be required to study this transition
definitively. 

Avid readers will have noticed that most of the individual test galaxies at
redshift $z = 0.5 - 1$ in Section \ref{sec:indicator_testindiv} are
less luminous than $10^{12}$ \Lsun. Space density of ULIRG at these
redshifts is too sparse for large number of them to be present in
pencil-beam surveys, such as those used in Section
\ref{sec:indicator_testindiv}. Therefore, while we have tested the new
indicator thoroughly with typical star-forming galaxies at $z = 0.5 -
1$, the formalism may not be adequate to describe luminous ULIRGs in
this redshift range. But these ULIRGs at the bright-end of the
luminosity function are often bright enough to be within reach of the
current far-IR, sub-mm, and radio facilities to provide direct
measurement of the \LIR. The results from the new 24 $\micron$
indicator should therefore be cross-checked with these independent
bands when possible.

We must also caution that there are very few galaxies with \LIR\ $>
10^{13}$ \Lsun\ in the seven tests in Section \ref{sec:indicator_test}
due to their rarity, even at $z > 2.5$. This hyperluminous regime
could be dominated by merger-triggered compact starbursts even at high
$z$ \citep[e.g., ][]{Hopkins10}, which will result in an
underestimation of \LIR\ and SFR by up to 0.5 dex (refer to
color-coded lines representing compact objects as a function of $z$ in
Figures \ref{compare_lir_CDFS} and \ref{compare_lir_HDFN}). For the
same reasons that prevent us from thoroughly testing this luminosity
regime, we expect compact hyperluminous starbursts to be rare, and
will not affect the usefulness of the new indicator within the goal of
providing a tracer of SFR in typical U/LIRGs (e.g., $L^*$ galaxies)
out to $z \sim 2.8$, where most of the stellar mass in the Universe
was formed. Again, these most luminous galaxies will be bright enough
for independent measurements of SFR using far-IR, sub-mm, and radio
observations.

\subsection{The Eddington Luminosity of
  ULIRGs}\label{sec:discuss_LEdd}   
It has been suggested that local ULIRGs are optically thick to mid-IR
photons and radiation pressure may play a role in limiting their
maximum luminosity \citep{Thompson05, Younger08, Thompson09}. If the
degree of compactness is assumed to be similar for both the local and
high-$z$ ULIRGs, the IR emission of high-$z$ ULIRGs with
\LIR\ commonly found far above their local counterparts might exceed
the Eddington luminosity. Our results ease the concerns about the
Eddington limit of ULIRGs at high $z$ for two reasons. First, galaxies
at higher \LIR\ are affected by the mid-IR excess to a greater
extent. The new indicator would therefore lower the \LIR\ estimates
especially for high-luminosity galaxies. To test the effects of
\LIR\ reduction to the luminosity functions of galaxies, we construct
a set of IR LFs of 24 $\micron$ sources at $0 < z < 2.5$ using the 24
$\micron$ observations from COSMOS \citep[e.g.,][]{Sanders07} and find
that the bright-ends of the LFs in all redshift bins are below
\LIR\ of $\sim 10^{13}$ \Lsun\ (quantitatively, galaxies with
\LIR\ above $10^{13}$ \Lsun\ are rarer than $10^{-6}/$Mpc$^3/$dex in
all redshift bins), which suggests that the maximum \LIR\ of
star-forming galaxies is in general $\lesssim 10^{13}$ \Lsun. Second,
the success of the corrections for the size evolution described in
\citet{Rujopakarn11} as well as this work to account for the general
IR SED evolution indicate that the great majority of high-$z$ IR
galaxies are physically extended, which lowers the optical depth of
the galaxy. Both the lower \LIR\ and optical depth imply that high-$z$
ULIRGs are emitting below the Eddington limit and that the maximum
\LIR\ of these galaxies is not governed by radiation pressure. \\

\section{CONCLUSIONS}\label{sec:conclusions}
The use of single-band mid-IR indicators (such as 24 $\micron$
observations) to estimate \LIR\ has previously been affected by the
differences between the structure of local U/LIRGs, upon which the SED
templates were constructed, and those of the galaxies in the
cosmological surveys being studied. Starbursts in the local U/LIRGs
are interaction-induced and compact (sub-kpc), while U/LIRGs at high
$z$ are typically much more extended ($\sim$ few kpc), with surface
areas $\sim 100\times$ larger than their local counterparts. The
resulting larger surface area of the photodissociation regions
harboring aromatic emission leads to stronger aromatic features at
high $z$ that appears as an evolution in aromatic strength with
$z$. The ultimate implication of this morphological difference is that
single-band mid-IR observations, which at $z > 1$ probe the
aromatic-dominated SED region, will overpredict \LIR\ and SFR compared
to far-IR measurements if the relationship of local SED templates to
\LIR\ is assumed for the calculation of the bolometric and
k$-$corrections.

Following the \citet{Rujopakarn11} result that \LIRSD\ can serve as a
good predictor for the appropriate SED of star-forming galaxies, we
construct a new 24 $\micron$ single-band \LIR\ and SFR indicator for
star-forming galaxies at redshift $0 < z < 2.8$. The resulting $0.03$ dex 
average agreement (0.13 dex scatter) between \LIR\ estimates from the
new single-band 24 $\micron$ indicator and those from multi-band
far-IR SED fitting support the \citet{Rujopakarn11} result that
\LIRSD\ is indeed the dominant factor affecting the SED shapes of
star-forming galaxies.

For the purpose of estimating \LIR\ and SFR, we recommend using the
new indicator at $0 < z < 2.8$. A step-by-step
recipe is given in Section \ref{sec:final_SFR}. The new indicator
converges with that of \citet{Rieke09} for local and low $z$ galaxies
(Section \ref{sec:discuss_sfmodes}). But unlike the \citet{Rieke09}
indicator, the new indicator carries the extended structure assumption
and could underestimate \LIR\ for the minority of cases ($10\%$) of
high-$z$ starbursts that are compact (Figures \ref{compare_lir_CDFS}
and \ref{compare_lir_HDFN}). Thus, if an object is known {\it a
  priori} to be a compact starburst (e.g., from high-resolution
optical or interferometric radio observations), we recommend readers
to apply the bolometric correction factor given by \citet{Rieke09} in
their Equation 14. On the other hand, where the structure of the
star-forming region is not known {\it a priori}, we quote an overall scatter
of 0.13 dex, when AGNs are identified and excluded from the sample
using X-ray and $3.6-8.0$ $\micron$ (i.e. {\it Spitzer}/IRAC)
observations. 

We thank Daniel Eisenstein for insightful discussions, and J. D. Smith
for the L$_{{\rm PAH}}/$\LIR\ data. This work is supported by contract
1255094 from Caltech/JPL to the University of Arizona. WR gratefully
acknowledges the support from the Thai Government Scholarship. 

\appendix

\section{Mitigation of Potentially Blended Objects in the {\it
    Herschel} Observations}\label{sec:rejectblended}
Since the {\it Herschel} PACS and SPIRE source extraction procedure is
designed to yield a sample for the fiducial \LIR\ estimates to test
the new indicator, an emphasis is given to achieving an unbiased
sample, rather than maximizing the number of detectable sources. To
this end, we develop a figure of merit (FoM) and criteria by which we
exclude potentially blended objects. This is important to mitigate
the cases of intervening objects contributing additional flux to the object of
interest. The FoM is defined in an unbiased manner based on the 24
$\micron$ prior catalog by 
\begin{equation}
\label{eq55}{\rm FoM} = \sum_i\left[({\rm
    R}_i/4)^{-1}(d_i/6'')^{-1}\right]\end{equation}
\noindent where R$_i$ is the ratio of the 24 $\micron$ flux of the
prior object to the 24 $\micron$ fluxes of other nearby objects within
the radius (9.1'') of SPIRE 250 $\micron$ beam from the prior and
$d_i$ is the distance from the prior object to the corresponding
nearby objects. The 250 $\micron$ beam is chosen for this criteria
because it is the band at which most of our ECDFS and HDFN samples are
detected. The normalizations for R$_i$ and $d_i$ are chosen such that
the FoM increases rapidly if other sources that are expected to be in
the beam could be in within the same SPIRE pixel (6'') or could have  
comparable SPIRE fluxes to the object at the prior position. The R$_i$
normalization value is determined by inspecting the variation of the
ratios of 24 and 250 $\micron$ fluxes for SED templates in the
\citet{Rieke09} library, which gives a $\sim$4$\times$ variation. That
is, if the 24 $\micron$ flux of a nearby object is less than $\sim
1/4$ that of the prior object, it is unlikely that its 250 $\micron$
flux will be comparable to the prior (and hence cause a blending
problem) at redshift $0 - 2.8$, assuming the \citet{Rieke09} SED
shape. We emphasize that the exclusion does not depend sensitively on
the normalizations in the FoM, and we have adopted a conservative
rejection threshold by excluding objects with FoM $> 5$ from the final
catalog, rejecting only the most likely blended objects.

\newcommand{\noopsort}[1]{} \newcommand{\printfirst}[2]{#1}
  \newcommand{\singleletter}[1]{#1} \newcommand{\switchargs}[2]{#2#1}

\clearpage
\begin{center}
\begin{deluxetable}{llcccccc}
\tablewidth{0pt}
\tablecaption{Local Galaxies with \LIRSD\ Measurement (Online Material)}
\tablehead{ID & {\it IRAS} ID & Distance\tablenotemark{a} & Diameter & \LIR\tablenotemark{a} &
  \LIRSD\ & $\sigma \Sigma_L$ & Ref.\tablenotemark{b} \\ \cline{6-7}
 & & Mpc & kpc & log(\Lsun) & \multicolumn{2}{c}{log(\Lsun/kpc$^2$)} & } 
\startdata
NGC2976 & F09431+6809 &   3.8 &  1.5 &  8.61 &  8.36 &  0.11 & 1 \\
NGC4826 & F12542+2157 &   6.0 &  1.2 &  9.14 &  9.09 &  0.11 & 1  \\
NGC2403 & F07320+6543 &   3.8 &  3.0 &  9.25 &  8.40 &  0.11 & 1  \\
NGC925 & F02242+3321 &   9.8 &  6.1 &  9.46 &  7.99 &  0.11 & 1  \\
NGC1512 & ... &  11.3 &  3.2 &  9.49 &  8.58 &  0.11 & 1  \\
NGC5866 & F15051+5557 &  13.0 &  1.8 &  9.51 &  9.10 &  0.11 & 1  \\
NGC2841 & ... &  10.5 &  4.9 &  9.52 &  8.24 &  0.11 & 1  \\
NGC4559 & F12334+2814 &  11.9 &  5.0 &  9.62 &  8.33 &  0.11 & 1  \\
NGC4736 & 12485+4123 &   5.7 &  1.7 &  9.79 &  9.43 &  0.11 & 1  \\
NGC3198 & F10168+4547 &  14.7 &  3.8 &  9.81 &  8.76 &  0.11 & 1  \\
NGC3184 & 10152+4140 &  11.9 &  6.5 &  9.86 &  8.34 &  0.11 & 1  \\
NGC3351 & F10413+1157 &  10.8 &  2.2 &  9.89 &  9.31 &  0.11 & 1  \\
NGC3938 & F11502+4423 &  13.1 &  4.7 &  9.99 &  8.75 &  0.11 & 1  \\
NGC4569 & F12343+1326 &  17.8 &  3.1 & 10.08 &  9.20 &  0.11 & 1  \\
NGC5055 & F13135+4217 &   8.4 &  4.3 & 10.15 &  8.99 &  0.11 & 1  \\
NGC5033 & F13111+3651 &  14.7 &  1.1 & 10.19 & 10.21 &  0.11 & 1  \\
NGC3627 & F11176+1315 &   9.3 &  3.5 & 10.44 &  9.46 &  0.11 & 1  \\
NGC5194 & F13277+4727 &   8.8 &  5.7 & 10.48 &  9.07 &  0.11 & 1  \\
NGC7331 & F22347+3409 &  16.2 &  4.7 & 10.64 &  9.40 &  0.11 & 1  \\
NGC23 & F00073+2538 &  63.9 &  1.2 & 11.11 & 11.06 &  0.11 & 2 \\
NGC6701 & F18425+6036 &  60.6 &  0.7 & 11.11 & 11.52 &  0.11 & 2  \\
UGC1845 & F02208+4744 &  66.4 &  0.8 & 11.13 & 11.42 &  0.11 & 2 \\
NGC5936 & F15276+1309 &  65.1 &  0.6 & 11.13 & 11.74 &  0.11 & 2  \\
MCG+02-20-003 & F07329+1149 &  72.4 &  0.8 & 11.14 & 11.46 &  0.11 & 2  \\
NGC2369 & F07160-6215 &  47.1 &  0.8 & 11.16 & 11.44 &  0.11 & 2  \\
ESO320-G030 & F11506-3851 &  40.4 &  0.9 & 11.16 & 11.35 &  0.11 & 2  \\
IC5179 & F22132-3705 &  50.0 &  1.6 & 11.22 & 10.92 &  0.11 & 2  \\
NGC2388 & F07256+3355 &  61.9 &  0.8 & 11.29 & 11.55 &  0.11 & 2  \\
NGC7771 & F23488+1949 &  61.2 &  1.0 & 11.40 & 11.47 &  0.11 & 2  \\
MCG+12-02-001 & F00506+7248 &  68.9 &  0.8 & 11.50 & 11.80 &  0.11 & 2  \\
... & F03359+1523 & 146.9 &  0.1 & 11.53 & 13.94 &  0.17 & 3 \\
NGC1614 & F04315-0840 &  67.1 &  0.9 & 11.66 & 11.87 &  0.11 & 4 \\
UGC2369 & F02512+1446 & 130.7 &  0.1 & 11.66 & 13.96 &  0.16 & 3 \\
Arp236 & F01053-1746 &  84.2 &  2.6 & 11.71 & 10.98 &  0.11 & 4 \\
Arp193 & F13182+3424 & 107.1 &  0.8 & 11.73 & 12.03 &  0.11 & 4 \\
UGC4881 & F09126+4432 & 172.7 &  0.1 & 11.75 & 13.56 &  0.12 & 3 \\
Arp299 & F11257+5850 &  51.2 &  0.4 & 11.94 & 12.80 &  0.11 & 4 \\
... & F17132+5313 & 218.9 &  0.1 & 11.95 & 13.76 &  0.15 & 3 \\
... & F15163+4255 & 180.8 &  0.2 & 11.95 & 13.29 &  0.11 & 3 \\
... & F10565+2448 & 188.9 &  0.8 & 11.99 & 12.29 &  0.11 & 4 \\
VIIZw31 & F05081+7936 & 230.2 &  2.1 & 12.00 & 11.46 &  0.11 & 5 \\
... & F23365+3604 & 269.8 &  1.1 & 12.19 & 12.21 &  0.11 & 5 \\
Arp220 & F15327+2340 &  85.6 &  0.2 & 12.27 & 13.77 &  0.11 & 6 \\
... & F17207-0014 & 188.2 &  0.9 & 12.45 & 12.65 &  0.10 & 4
\enddata
\tablenotetext{a}{Distance and \LIR\ from \citet{Sanders03} and
  adjust to match our cosmology.}
\tablenotetext{b}{Diameters references 1. measured from {\it Spitzer}
  MIPS 24 $\micron$ imaging taken by SINGS
  \citep[e.g.,]{Calzetti07}, 2. measured from {\it
    Hubble} NICMOS Pa-$\alpha$ taken by \citet{AH06}, 3. 8.4 GHz
  radio sizes given by \citet{Condon91}, 4. CO ($3-2$) sizes given by
  \citet{Iono09}, 5. CO ($2-1$) or CO ($1-0$) sizes given by
  \citet{DS98}, 6. 5 GHz radio size based on \citet{Rovilos03}.}
\tablecomments{Reproduced from Table 2 of
  \citet{Rujopakarn11}. Uncertainties in \LIR\ estimates are dominated
  by distance measurement uncertainties, which are not published in
  the \citet{Sanders03} catalog; we use error bars of 0.1 dex for
  \LIR-\LIRSD\ parameterizations fitting and to estimate the $\sigma
  \Sigma_L$.} 
\label{table_4}
\end{deluxetable}
\end{center}

\begin{center}
\begin{deluxetable}{lccccccc}
\tablewidth{0pt}
\tablecaption{High-$z$ Galaxies with \LIRSD\ Measurement (Online Material)}
\tablehead{ID & $z$ & Diameter & \LIR$_{\rm old}$ & $\sigma L$ & \LIRSD\ & $\sigma \Sigma_L$  & Ref.\tablenotemark{a} \\ \cline{4-5}
  & & kpc & \multicolumn{2}{c}{log(\Lsun)} & \multicolumn{2}{c}{log(\Lsun/kpc$^2$)} &  } 
\startdata
RGJ105146.61+572033.4 & 2.383 &  4.2 & 13.37 & 0.135 & 12.23 & 0.247 & 3 \\
SMMJ105151.69+572636.1 & 1.147 &  6.1 & 12.60 & 0.139 & 11.14 & 0.209 & 4, 5 \\
RGJ105154.19+572414.6 & 0.922 &  4.0 & 12.21 & 0.132 & 11.11 & 0.253 & 3 \\
SMMJ105155.47+572312.8 & 2.686 &  2.2 & 13.13 & 0.157 & 12.55 & 0.372 & 4, 5\\
SMMJ105158.02+571800.3 & 2.239 &  6.7 & 13.21 & 0.146 & 11.67 & 0.203 & 4, 5\\
RGJ105159.90+571802.4 & 1.047 &  5.4 & 12.83 & 0.132 & 11.47 & 0.213 & 3 \\
SMMJ105201.25+572445.8 & 2.148 &  3.3 & 12.90 & 0.138 & 11.97 & 0.295 & 4, 5 \\
J123607+621328 & 0.435 &  5.3 & 10.84 & 0.131 &  9.50 & 0.164 & 6, 7, 8 \\
J123615+620946 & 1.263 &  5.2 & 12.16 & 0.133 & 10.84 & 0.194 & 6, 7, 8 \\
SMMJ123616.15+621513.7 & 2.578 &  8.0 & 13.72 & 0.131 & 12.02 & 0.176 & 2, 6, 7, 8 \\
J123617+621011 & 0.845 &  3.6 & 11.05 & 0.135 & 10.04 & 0.229 & 6, 7, 8 \\
J123618+621550 & 1.870 &  2.6 & 12.96 & 0.131 & 12.24 & 0.181 & 6, 7, 8 \\
J123619+621252 & 0.473 &  1.8 & 11.62 & 0.130 & 11.22 & 0.183 & 6, 7, 8 \\
SMMJ123622.65+621629.7 & 2.466 &  9.7 & 13.72 & 0.130 & 11.85 & 0.160 & 2, 6, 7, 8 \\
SMMJ123629.13+621045.8 & 1.013 &  6.6 & 12.71 & 0.130 & 11.18 & 0.143 & 1, 2, 6 \\
J123630+620923 & 0.953 &  3.7 & 11.80 & 0.131 & 10.77 & 0.223 & 6, 7, 8 \\
J123633+621005 & 1.016 &  7.5 & 12.58 & 0.130 & 10.94 & 0.165 & 6, 7, 8 \\
J123634+621213 & 0.456 &  5.4 & 11.72 & 0.130 & 10.36 & 0.144 & 6, 7, 8 \\
J123634+621241 & 1.219 &  6.4 & 13.08 & 0.130 & 11.58 & 0.138 & 6, 7, 8 \\
J123635+621424 & 2.011 &  1.9 & 14.03 & 0.130 & 13.58 & 0.195 & 6, 7, 8 \\
J123641+620948 & 0.518 &  2.9 & 11.33 & 0.130 & 10.51 & 0.173 & 6, 7, 8 \\
RGJ123645.89+620754.1 & 1.433 &  4.2 & 13.36 & 0.131 & 12.22 & 0.193 & 3, 6 \\
J123646+620833 & 0.971 &  4.9 & 12.76 & 0.130 & 11.49 & 0.192 & 6, 7, 8 \\
J123649+621313 & 0.475 &  4.6 & 11.14 & 0.131 &  9.92 & 0.179 & 6, 7, 8 \\
J123650+620801 & 0.559 &  3.5 & 11.09 & 0.131 & 10.11 & 0.213 & 6, 7, 8 \\
J123651+621030 & 0.410 &  5.1 & 11.44 & 0.130 & 10.13 & 0.165 & 6, 7, 8 \\
RGJ123653.37+621139.6 & 1.275 &  5.8 & 13.02 & 0.131 & 11.60 & 0.168 & 3, 6 \\
J123655+620917 & 0.419 &  2.6 & 11.39 & 0.130 & 10.66 & 0.181 & 6, 7, 8 \\
J123655+620808 & 0.792 &  4.1 & 12.22 & 0.130 & 11.09 & 0.167 & 6, 7, 8 \\
J123659+621449 & 0.761 &  5.7 & 11.84 & 0.130 & 10.43 & 0.174 & 6, 7, 8 \\
J123704+620755 & 1.253 &  6.5 & 13.26 & 0.131 & 11.74 & 0.174 & 6, 7, 8 \\
J123705+621153 & 0.902 &  9.1 & 12.31 & 0.130 & 10.49 & 0.152 & 6, 7, 8 \\
J123708+621056 & 0.422 &  3.0 & 11.26 & 0.130 & 10.42 & 0.211 & 6, 7, 8 \\
RGJ123710.60+622234.6 & 1.522 &  3.4 & 12.98 & 0.171 & 12.02 & 0.237 & 3 \\
SMMJ123711.98+621325.7 & 1.992 &  6.9 & 12.79 & 0.132 & 11.22 & 0.197 & 1, 2, 6 \\
J123713+621603 & 0.938 &  5.5 & 11.72 & 0.131 & 10.34 & 0.183 & 6, 7, 8 \\
J123714+621558 & 0.567 &  7.6 & 10.93 & 0.131 &  9.27 & 0.154 & 6, 8 \\
J123716+621643 & 0.557 &  4.5 & 11.51 & 0.130 & 10.31 & 0.187 & 6, 7, 8 \\
J123716+621007 & 0.411 &  3.0 & 11.18 & 0.130 & 10.33 & 0.211 & 6, 7, 8 \\
J123721+621346 & 1.019 &  4.4 & 11.97 & 0.131 & 10.79 & 0.210 & 6, 7, 8
\enddata
\tablenotetext{a}{1. \citet{Chapman04}, 2. \citet{Chapman05},
  3. \citet{Casey09}, 4. \citet{BiggsIvison08}, 5 \citet{Ivison07}
  6. GOODS {\it Spitzer} Legacy Data, Dickinson
  et al., in prep., 7. \citet{Muxlow05},
  8. \citet{Morrison10}}
\tablecomments{Reproduced from Table 1 of \citet{Rujopakarn11} with
  omission of objects whose diameters were measured with CO
  observations, which are excluded from the fit in Section
  \ref{sec:indicator}.} 
\label{table_5}
\end{deluxetable}
\end{center}

\end{document}